\documentclass{aa}
\usepackage{txfonts}
\usepackage{graphicx}
\usepackage{natbib}
\bibpunct{(}{)}{;}{a}{}{,}

\begin{document}
\title{New $\lambda$6\ cm observations of the Cygnus Loop}
\author{ X. H. Sun\inst{1} 
         \and W. Reich\inst{2} 
         \and J. L. Han\inst{1} 
         \and P. Reich\inst{2} 
         \and R. Wielebinski\inst{2}}
\offprints{W. Reich}
\institute{ National Astronomical Observatories, Chinese Academy of 
            Sciences, Beijing 100012, China\\
            \email{xhsun,hjl@bao.ac.cn}
            \and Max-Planck-Institut f\"{u}r Radioastronomie, Auf dem H\"ugel 69, Bonn 53121, Germany\\
            \email{wreich,preich,rwielebinski@mpifr-bonn.mpg.de}}
\date{Received / Accepted}
\abstract{Radio continuum and polarization observations of the entire Cygnus
Loop at $\lambda$6~cm wavelength were made with the Urumqi 25~m telescope.
The $\lambda$6~cm map is analysed together with recently published
maps from the Effelsberg 100~m telescope at $\lambda$21~cm and
$\lambda$11~cm. The integrated flux density of the Cygnus Loop at
$\lambda$6~cm is 90$\pm$9~Jy, which implies a spectral index of
$\alpha=-0.40 \pm 0.06$ being consistent with that
of \citet{ury+04} in the wavelength range up to $\lambda$11~cm. 
This rules out any global spectral steepening up to $\lambda$6~cm. 
However, small spectral index variations in
some regions of the source are possible, but there are no indications for
any spectral curvature.  The linear polarization data at $\lambda$6~cm show
percentage polarizations up to 35\% in some areas of the Cygnus Loop,
exceeding those observed at $\lambda$11~cm. The Rotation Measure is around
$-21$~rad m$^{-2}$ in the southern area, which agrees with previous observations.
However, the distribution of Rotation Measures is rather complex in the
northern part of the Cygnus Loop, 
where the $\lambda$21~cm emission is totally depolarized.
Rotation Measures based on $\lambda$11~cm
and $\lambda$6~cm data are significantly larger than in the southern
part. The difference in the polarization characteristic between the northern and
southern part supports previous ideas 
that the Cygnus Loop consists of two supernova remnants.

\keywords{ISM: magnetic fields -- supernova remnants -- radio continuum -- 
polarization}}
\titlerunning{Cygnus Loop at $\lambda$6\ cm}
\maketitle

\section{Introduction}
The Cygnus Loop (G74.0$-$8.5) is a large and intense non-thermal Galactic radio 
source. It is a rather well studied object at all observing bands. 
It has an apparent
size of $\sim 4\degr \times 3\degr$ and its distance was recently 
revised to 540\ pc by \citet{bsr05} using {\it Hubble Space Telescope} data. 
The small distance and its location well out of the Galactic plane 
lead to little obscuration or confusion with 
Galactic emission to any observations. Its non-thermal radio spectrum makes its 
identification as a supernova remnant (SNR) most likely, which is supported by 
observations in other wavelength ranges.

In the radio band a recent spectral index study of the Cygnus Loop
was made by \citet{ury+04} 
using high quality data from the Effelsberg 100\ m telescope and the DRAO 
synthesis telescope in the frequency range from 408\ MHz to 2675\ MHz at arcmin angular 
resolution. For the integrated radio emission they obtained a spectral index of 
$\alpha=-0.42\pm0.06$ (with $\alpha$ defined as $S_\nu\propto\nu^\alpha$, with $S_\nu$ 
being the flux density and $\nu$ the frequency). This straight spectrum rules out any 
spectral break around 1\ GHz as reported earlier \citep{den74}. However, the flux 
density measured by \citet{kb72} at 5\ GHz, the highest frequency where the Cygnus 
Loop was observed so far, is inconsistent with the spectrum of 
\citet{ury+04} and indicates some spectral steepening above 2.7~GHz. A steepening in the 
spectrum of a SNR is an important and 
characteristic feature, which is closely connected to the age and the evolution
 of the source in the interstellar medium. This is for instance observed
for the SNR S147, which is similar to the Cygnus Loop in many 
aspects \citep{fr86}. It is therefore of interest to establish the spectral 
break for the Cygnus Loop and to identify regions or structures within the SNR
 where the spectral break takes place. However, sensitive radio continuum observations 
of large sources like the Cygnus Loop are not easy to perform. Only single-dish
telescopes can do that at high frequencies. However, they need a high 
sensitivity and baseline stability of the receiving system (a low 1/f-noise) so 
that any large-scale emission component is detected, otherwise it will lead to an 
underestimate of the integrated flux density and falsely suggest a steepening of
the spectrum. 

The Cygnus Loop is not fully understood, though it has been observed in almost all 
bands of the electromagnetic spectrum. In particular its morphology with a large 
northern circular shell and a bubble-like southern part makes it an unusual object 
compared to other SNRs. It was argued \citep{al99} that the southern bubble 
resembles the outbreak of the SNR as visible in the X-ray bright northern shell into 
a low density cavity of the interstellar medium. However, theoretical modeling by
\citet{ten85} assuming a molecular environment shows that the SNR must be 
exploded in the southern bubble and the northern part is the outbreak.
Substantial differences between the northern part and the southern 
part have been observed in their radio emission properties~\citep{ury+04} as well as in 
their optical and X-ray emission characteristics~\citep{pfr+02}. This 
motivated \citet{ury+02} to propose that the Cygnus Loop consists of two 
likely interacting SNRs: G74.3$-$8.4 and G72.9$-$9.0. That the Cygnus Loop consists
of two SNRs was independently put forward by \citet{lea02}. However, more data are
needed to settle this view.

In this paper we try to resolve these questions using sensitive $\lambda$6\ cm 
observations of this large radio source made with a new $\lambda$6\ cm receiving system 
installed at the Urumqi 25\ m telescope. We describe the receiving system and the data 
reduction procedure in some detail in Section 2. Our results and the analysis of the 
total intensity and polarized emission of the Cygnus Loop at 
$\lambda\lambda$6\ cm, 11\ cm and 21\ cm are presented in Section 3, followed by 
remarks on the two--SNR scenario and the conclusion in Section 4 and 
5, respectively.

\section{Observations and Data Reduction}

The $\lambda$6\ cm observations were made with the 25\ m telescope at Nanshan station 
operated by the Urumqi Astronomical Observatory, which is part of the National 
Astronomical Observatories of the Chinese Academy of Sciences. The telescope is 
located about 70\ km south of Urumqi city at an altitude of 2029\ m above sea 
level with the geographic longitude of 87\degr E and latitude of $+$43\degr. The 
telescope was mainly used for VLBI observations as a 
member of the European VLBI Network and also for pulsar timing observations 
at L-band \citep{wang+01,wang+03}. 
New possibilities for continuum and polarization mapping were 
opened by the recent installation of a dual-channel $\lambda$6\ cm receiving system 
constructed at the Max-Planck-Institut f\"ur Radioastronomie (MPIfR) 
in Bonn/Germany. This receiver is a copy of the $\lambda$6\ cm receiver 
being in operation at the MPIfR Effelsberg 100\ m telescope since 1996. 
The new receiver has a higher stability and a lower 
1/f-noise, which are ideal characteristics to perform mapping of 
large areas of the sky in the radio continuum and in linear polarization. 

The $\lambda$6\ cm receiver is installed at the secondary focus of the 25\ m 
telescope. Following the corrugated circular feed horn an orthogonal transducer 
converts the signal into left-hand (L) and right-hand (R) circularly polarized components, 
which are then amplified by two cooled HEMT pre-amplifiers working below 15\ K. 
A ``Digital Backend'' from the MPIfR collects the data at a sampling 
rate of 32\ msec. Every 32\ msec the frontend setting changes, so that either a 
calibration signal of 1.7\ K $\rm T_{a}$ is added to the antenna signal 
and/or the signal phase is switched off by $180\degr$. Four
phases include all possible combinations in 4$\times$32\ msec. This fast switching ensures 
a continuous gain control of the receiving system and the $180\degr$
phase switch allows to compensate for the quadratic terms of the 
IF-polarimeter, which is the same as those used at the Effelsberg 100\ m 
telescope for broadband polarization observations \citep{rw+02}. 

A computer with a LINUX operation system is used to store the raw 
data in the MBFITS format as developed for the Effelsberg 100\ m, the IRAM 30\ m 
and the APEX 12\ m telescopes (Hatchell 2003).
The raw data are further processed by the TOOLBOX-software package adapted from 
the Effelsberg 100\ m telescope. Via TOOLBOX, the raw data of the four backend 
channels (RR*, LL*, RL*, LR*) are converted from a time series into a tabulated 
format with a user 
specified fixed spatial separation on the sky using a sinc-interpolation 
function. The calibration signals of each subscan (row or column) of a map 
are extracted and fitted to account for any gain drifting and also to control the 
phase stability of the IF polarimeter. The tabulated maps are then transformed 
into NOD2 maps~\citep{has74} with the polarization U and Q-channels corrected
for the parallactic angles. The LINUX PC is also used to 
command the telescope control computer to move the telescope according to the 
mapping requirements. Mapping can be done in a number of different astronomical
coordinate systems and in the az/el coordinate system of the telescope. 

\begin{table}
   \caption{Observational parameters for the Cygnus Loop}
   \label{obspara}
  {\begin{tabular}{ll}\hline
  Frequency [GHz] & 4.8 \\
  Bandwidth [MHz] & 600 \\
  T$_{\rm sys}$ [K] & 22 \\
  HPBW [\arcmin ] & 9.5 \\
  Scan mode & RA and DEC \\
  Scan-Velocity [\degr /min] & 2 \\
  Scan Separation [\arcmin ] & 4 \\
  Map Size [$\degr \times \degr$] & 4.2$\times$4.8  \\
  Coverages I (PI) map & 5 (6)   \\
  RMS-I [mK] & 1.0 \\
  RMS-PI [mK] & 0.4 \\
  Observation Date & August-December 2004  \\ \hline
  Primary Calibrator & 3C286 \\
  Flux Density [Jy] & 7.5 \\
  Polarization Percentage [\%] & 11.3  \\
  Polarization Angle [$\degr$] & 33 \\ \hline
  \end{tabular}}   
\end{table}

From test observations made in August 2004 the relevant system and antenna 
parameters were determined. A detailed description of the system performance 
will be given elsewhere. In brief: the system 
temperature was measured to be about 22\ K for clear sky conditions towards 
the zenith. The beam is circular and the half power beam width (HPBW) was found 
to be 9\farcm5. The aperture efficiency was measured to be about 62\% and 
the beam efficiency is about 67\%. The maximum of the first sidelobes along the 
four feed support legs is about -17\ dB or 2\%. The conversion factor between 
Jy/beam area and the main beam brightness temperature is 
$\rm T_{\mathrm B}[{\mathrm K}]/S[{\mathrm Jy}] = 0.164$. The 
central frequency of the receiver is 4.8\ GHz and its bandwidth is 600\ MHz (see Table 1).

We made a number of $4{\fdg}2\times4{\fdg}8$ raster-scan maps of the Cygnus 
Loop area centered at $\alpha_{2000}=20^{\mathrm h}52^{\mathrm m}$, 
$\delta_{2000}=30\degr30\arcmin$ by scanning along right ascension or 
declination direction, respectively. All these observations were done in autumn~2004 
at clear sky. To avoid any contamination by the far--sidelobe 
response of the solar emission all the observations were conducted in the late 
evening or during the night. The scan velocity was always $2\degr$/min.  
The scan spacing of $4\arcmin$ provides full sampling. This means 2\ seconds 
of integration time for each pixel and about 153\ minutes for one coverage. 
The tracking and pointing errors for each coverage are within $1\arcmin$ in 
right ascension and $30\arcsec$ in declination as found by fitting strong 
compact sources within the field and comparing with positions provided by the 
NVSS \citep{ccg+98}.

The NOD2 based data reduction package for radio continuum and polarization 
observations from the Effelsberg 100\ m telescope was converted from the SOLARIS
operation system to the LINUX system for Urumqi observations. The individual 
maps were edited to remove spiky interference, to correct baseline curvatures 
by polynomial fitting, and to suppress scanning effects by applying an 
``unsharp masking method''~\citep{sr79}. All individual maps were on a relative 
baselevel, where the two ends of each subscan are set to zero. All maps 
observed in the two scanning directions were then ``weaved'' together by 
applying the method of~\citet{eg88}, which is quite powerful to destripe a set 
of maps in the Fourier domain.  

The flux density scale for the continuum and polarization observations and also the 
polarization angle was adapted to 3C\ 286 as the main calibrator (see Table~1). 
3C\ 48 and 3C\ 138 were
observed as secondary calibration sources, which were always visible during the 
Cygnus Loop observations. From the scatter of the measured calibration sources 
we quote an accuracy of our total intensity and polarized intensity scales of
about 5\%. The polarization angle is stable within $\pm 1\degr$.

\section{Result and analysis}

\subsection{Urumqi $\lambda$6\ cm map}

\begin{figure}
\centering
\includegraphics[bb=52 133 625 648,width=0.55\textwidth,clip,angle=-90]{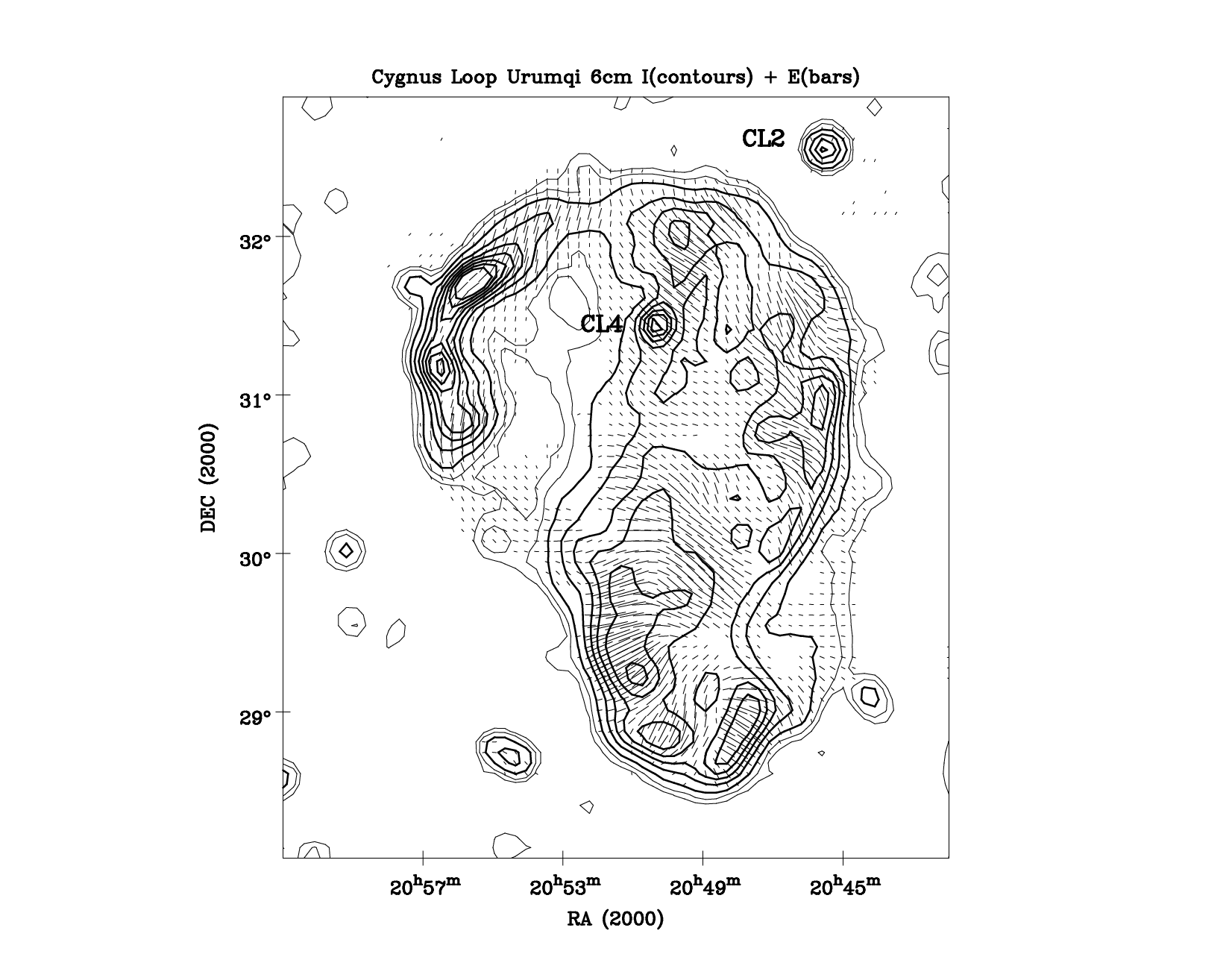}
\includegraphics[bb=52 133 625 648,width=0.55\textwidth,clip,angle=-90]{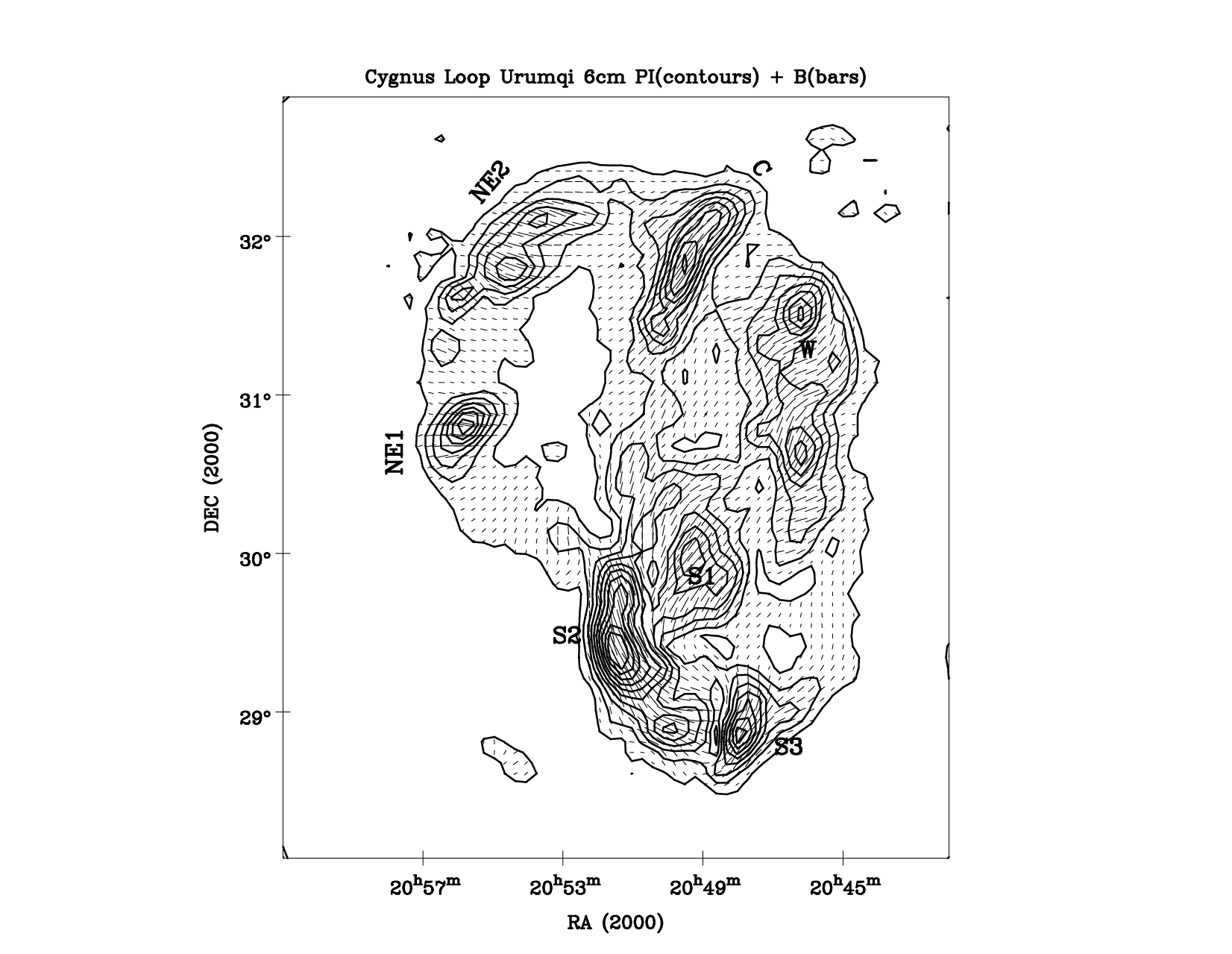}
   \caption{The Urumqi $\lambda$6\ cm map of the Cygnus Loop smoothed to a HPBW of 10\arcmin. 
Upper panel: Total intensity is displayed in 
contours, where the intensity levels of 5, 10 (thin lines), 20, 40, 
60 mK T$_{\rm B}$... (thick lines) are shown. 
The bars show the orientation of the polarization E-vectors with their length 
proportional to the polarized intensity and with the lower limit of 
2\ mK\ T$_{\mathrm B}$ (5$\times$ r.m.s-noise). The sources CL 2 and CL 4 are 
marked.
Lower panel: Contours show polarized intensities starting from  
2\ mK\ T$_{\mathrm B}$ and running in 3\ mK\ T$_{\mathrm B}$ steps. 
The bars are the same as in the upper panel but show the orientation of the 
B-vectors by adding $90\degr$ (this assumes negligible Faraday rotation). 
The main polarized features are labeled.}\label{6cm}
\end{figure}

From five full coverages of the Cygnus Loop and another partial one of 
its northern part we obtained the total intensity map as shown in the upper 
panel of Fig.~\ref{6cm}. For the maps in Stokes U and Q we have one full coverage 
in addition. From the U and Q maps we then calculated the polarized intensity map 
corrected for the noise level and the distribution of polarization angles.
The resultant r.m.s.-noise in our final total intensity map is about 
1\ mK\ T$_{\mathrm B}$ and  0.3\ mK\ T$_{\mathrm B}$ in the Stokes U and Q maps and 
about 0.4\ mK\ T$_{\mathrm B}$ in the polarized intensity map. These r.m.s-values 
are slightly larger than those expected from test observations of small 
fields, which can be attributed to "scanning effects" caused by instrumental 
drift, atmosphere and ground radiation variations that affect total 
intensity signals more severely than correlated polarization data. In addition, low-level 
interference close to the noise level can not be easily identified, but 
may also lower the sensitivity. A Gaussian 
fit to strong sources in the map yield an effective angular resolution of 
9\farcm7, also slightly larger than test observations reflecting small 
pointing differences between different coverages. All relevant
observational parameters are listed in Table~1.
The final results of our $\lambda$6\ cm observation are shown in 
Fig.~\ref{6cm} slightly smoothed to a HPBW of 10\arcmin. 
 
\begin{figure}[!htbp]
    \includegraphics[width=47mm,angle=-90]{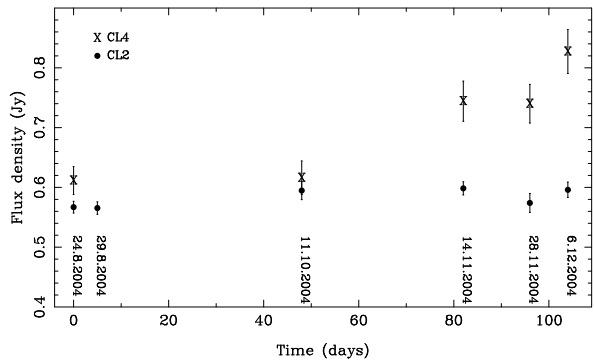}
    \caption{The flux densities of CL 2 and CL 4 are plotted versus 
             time. No flux density of CL 4 was obtained on
             August 29, 2004.}\label{cl4}
\end{figure}

During our observations we noted that the flux density of the source 
CL 4 (Fig.~\ref{6cm} upper panel, $\alpha_{2000}$: $20^{\rm h}50^{\rm m}46\fs3$, 
$\delta_{2000}$: $31\degr27\arcmin51\farcs0$) \citep{kwh+73} varied by about 30\% 
during a timescale of about 48 days~(Fig.~\ref{cl4}). Attention to the 
variability of CL 4 was first drawn by \citet{kwh+73}. Its location 
in the Cygnus Loop direction makes it a source of special interest. Meanwhile 
CL 4 was proved to be of extragalactic origin from H{\scriptsize I} 
absorption observation \citep{goss+79} with a redshift of 3.18~\citep{df01}. 
The observed variations cannot be explained by any instrumental effects 
since the simultaneous observation of CL 2 
(Fig.~\ref{6cm} upper panel, $\alpha_{2000}$: $20^{\rm h}45^{\rm m}44\fs2$,
$\delta_{2000}$: $32\degr33\arcmin42\farcs5$) did not show any 
variation (Fig.~\ref{cl4}). Apart its high variability CL 4 shows angular 
broadening of the turbulent interstellar medium as discussed by 
\citet{df01}. The flux density of 
0.58\ Jy for CL 2 is consistent with the spectral index of $\alpha$=$-$0.48 
obtained by~\citet{kwh+73} between 408\ MHz and 2695\ MHz.

\subsection{Effelsberg maps at $\lambda$21\ cm and $\lambda$11\ cm}
\begin{figure}[!htbp]
\centering
\includegraphics[bb=52 133 625 648,width=0.55\textwidth,clip,angle=-90]{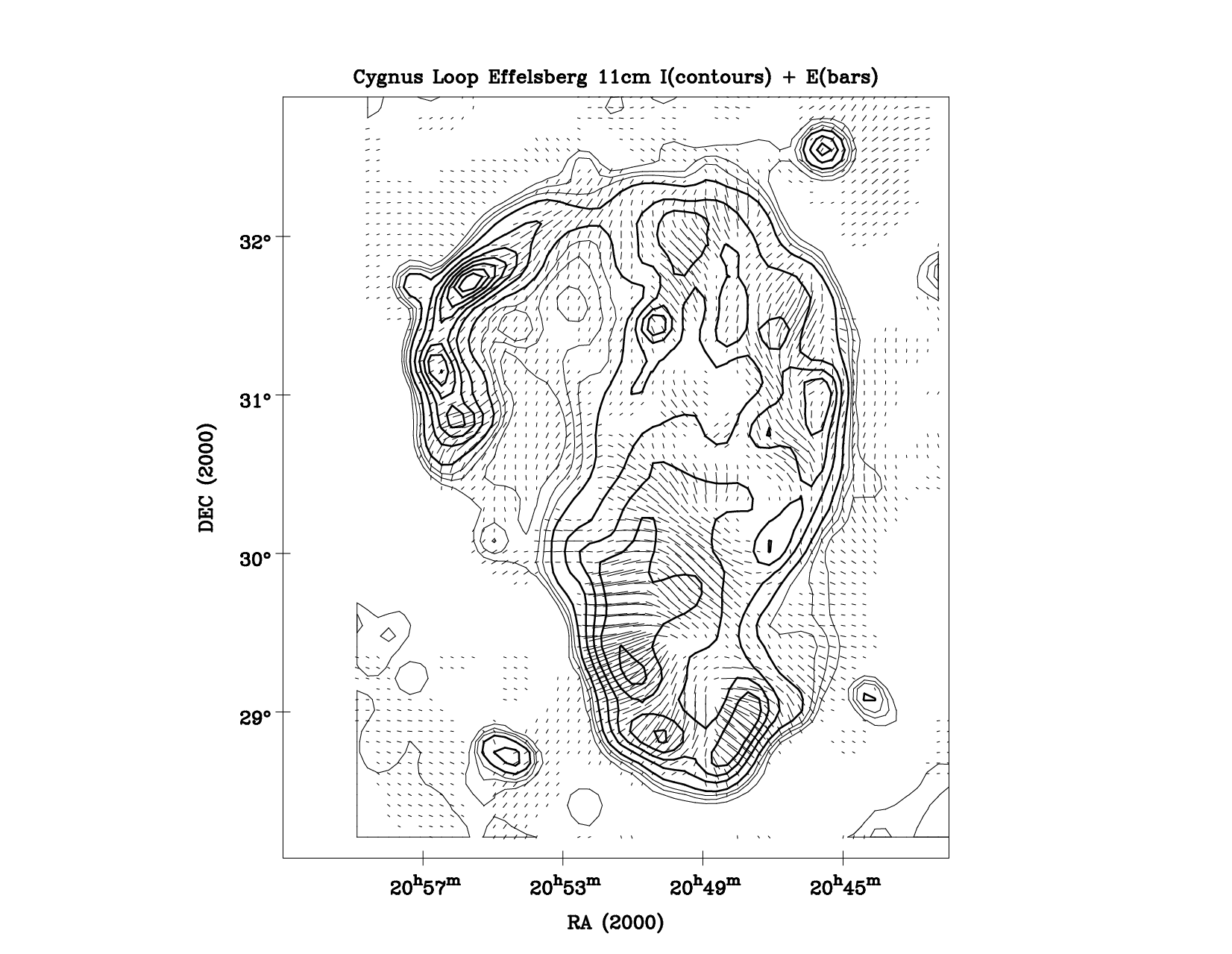}
\includegraphics[bb=52 133 625 648,width=0.55\textwidth,clip,angle=-90]{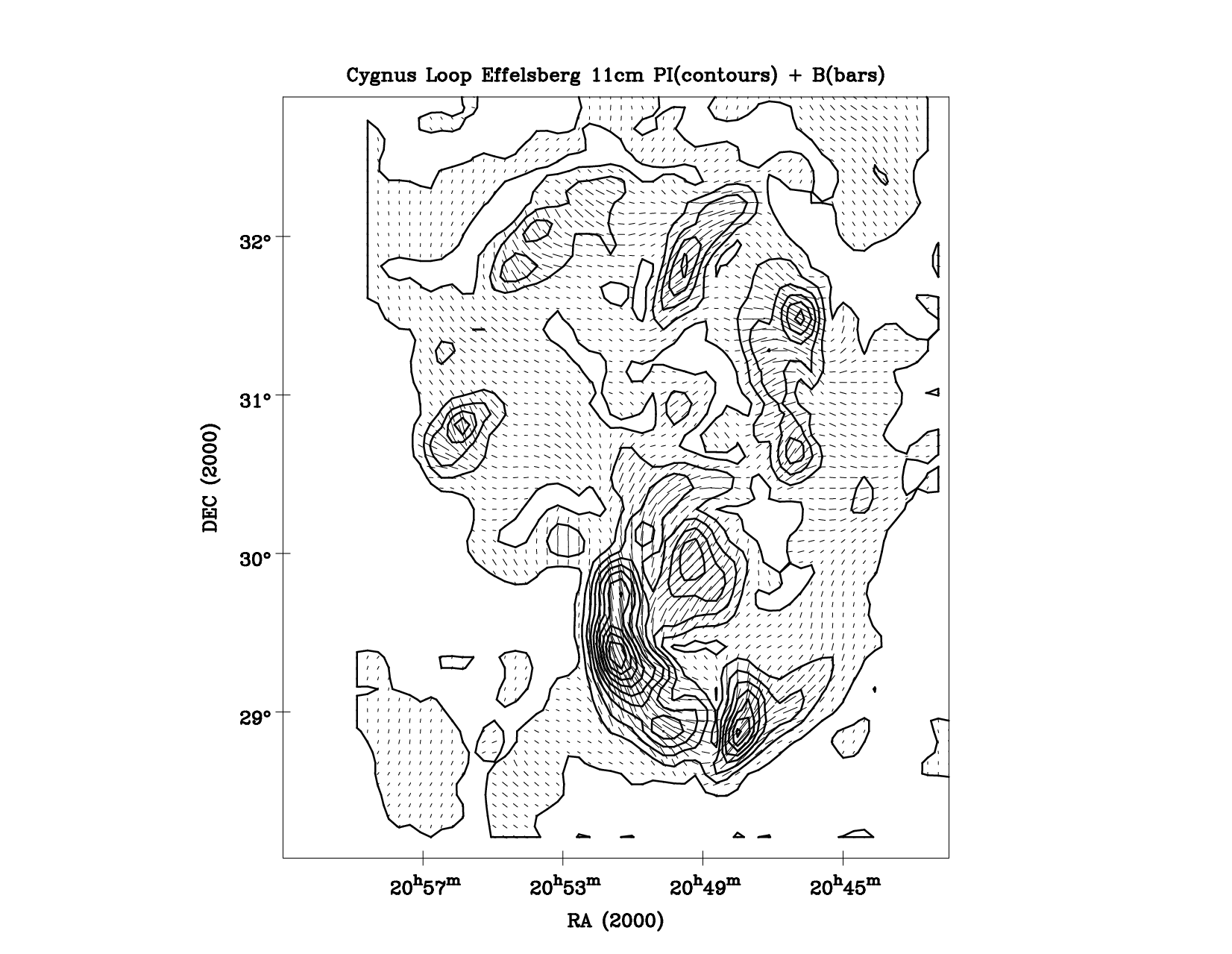}
      \caption{The same as Fig.~\ref{6cm} but for the Effelsberg $\lambda$11\ cm 
               maps. In the upper panel 
               the thin contours start from 20\ mK\ T$_{\mathrm B}$ 
               with an interval of 20\ mK\ T$_{\mathrm B}$. 
               The thick contours start at 
               100\ mK\ T$_{\mathrm B}$ and run in 100\ mK\ T$_{\mathrm B}$ 
               steps of total intensity. 
               In the lower panel, 
               the polarization intensity contours start at 7\ mK\ T$_{\mathrm B}$ and run in 
               14\ mK\ T$_{\mathrm B}$ steps.  
               Bars with polarization intensities
               below 7\ mK\ T$_{\mathrm B}$ (5$\times$ r.m.s-noise) are not shown. 
              }\label{11cm}
\end{figure}

\begin{figure}[!htbp]
\centering
\includegraphics[bb=52 133 625 648,width=0.55\textwidth,clip,angle=-90]{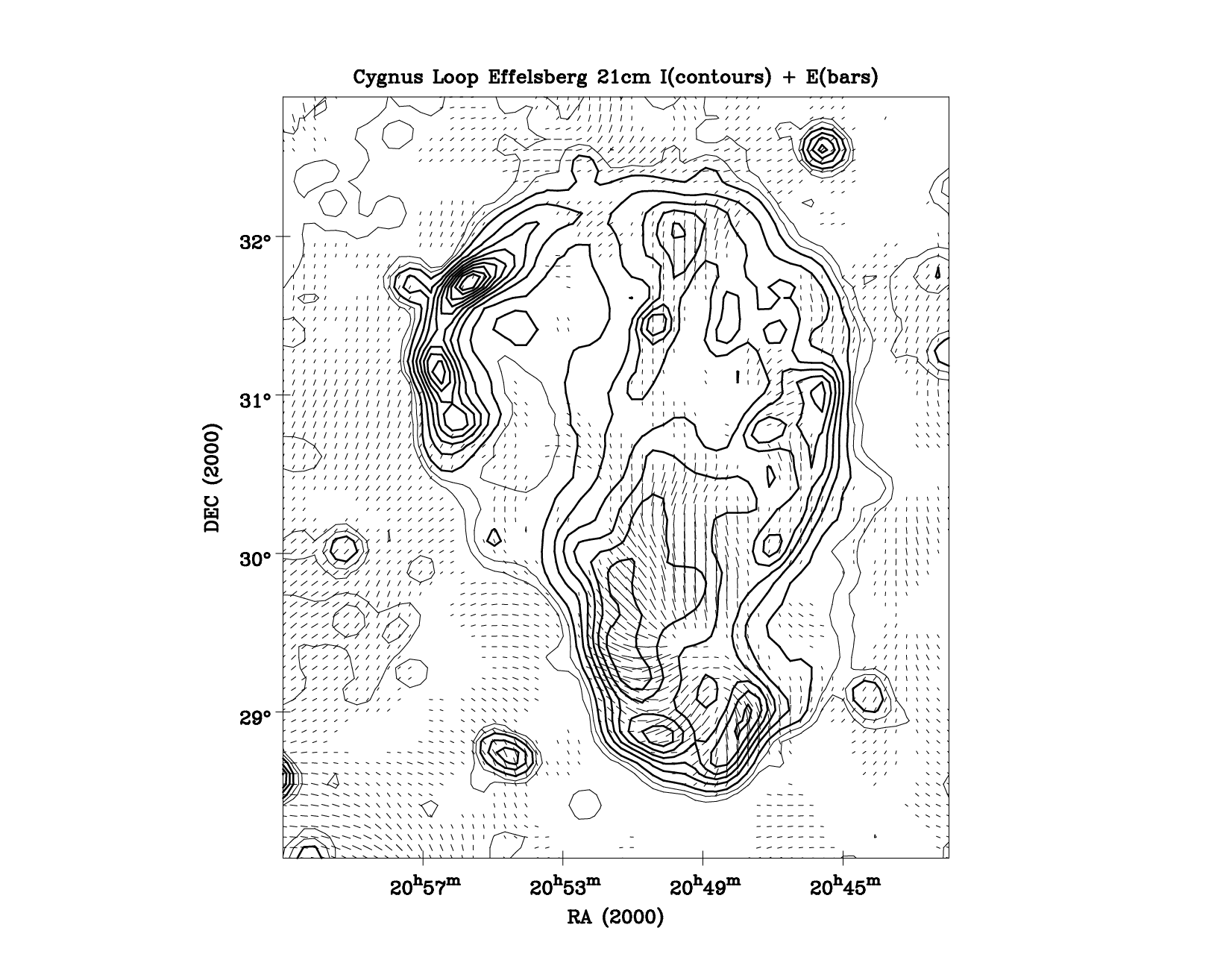}
\includegraphics[bb=52 133 625 648,width=0.55\textwidth,clip,angle=-90]{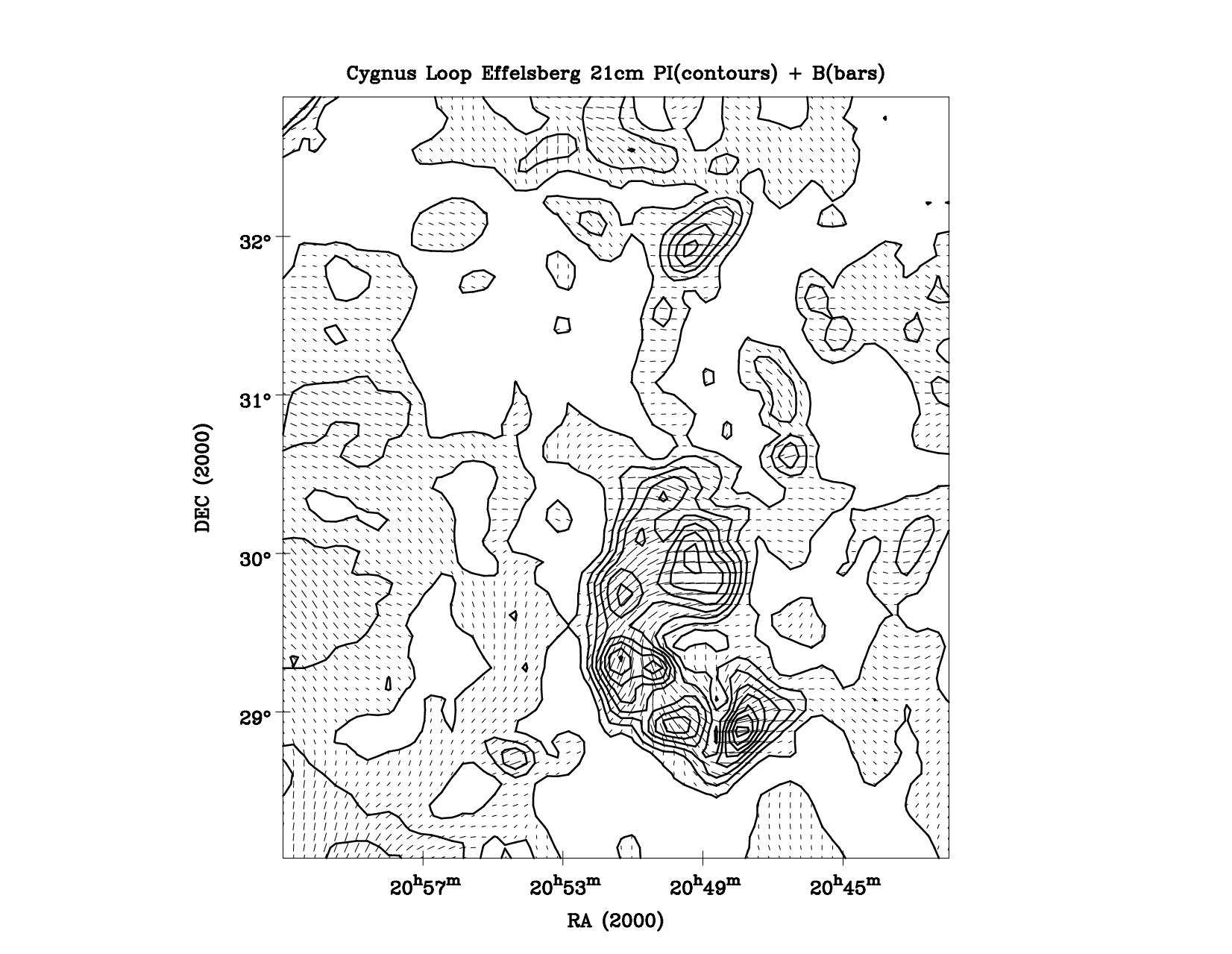}
      \caption{The same as Fig.~\ref{11cm} but for the $\lambda$21\ cm maps. 
               In the upper panel contours represent the total 
               intensities.  
               The thin contours are at 65 and 195\ mK\ T$_{\mathrm B}$
	       and the thick contours at 400, 800, 1200\ mK\ T$_{\mathrm B}$. 
               In the lower panel polarization intensity contours at 
               50, 100, 150\ mK\ T$_{\mathrm B}$ ... are shown. 
               Bars with polarization intensity smaller than 
               50\ mK\ T$_{\mathrm B}$ (5$\times$ r.m.s-noise) are not shown.}
       \label{21cm}
\end{figure}

In the following we compare our new $\lambda$6\ cm Cygnus Loop map with those at 
$\lambda$21\ cm and $\lambda$11\ cm from the Effelsberg 100\ m telescope. 
The $\lambda$21\ cm data were published by \citet{bu+99} as example maps for 
the Effelsberg ``Medium Galactic Latitude Survey'', where the Cygnus Loop area
was cut out from a larger field. It should be noted, that the polarization 
data were meanwhile reprocessed to better account for instrumental cross
talk effects. The differences to the published maps are small in general but 
only show up in areas of strong continuum sources. The $\lambda$11\ cm map 
was published by \citet{ury+02}. Both Effelsberg maps were convolved to a 
common angular resolution of $10\arcmin$ to be compared with the Urumqi map. 
At this angular resolution we measure a noise 
level of 13~mK and 10~mK at $\lambda$21\ cm and 4~mK and 1.4~mK at 
$\lambda$11\ cm for total intensity and polarized intensity, respectively. 
The smoothed Effelsberg $\lambda$11\ cm and $\lambda$21\ cm maps  
 are shown in Fig.~\ref{11cm} and Fig.~\ref{21cm}, respectively. 


\subsection{The integrated flux density spectrum of the Cygnus Loop}

\begin{figure}[!htbp]
    \includegraphics[width=0.3\textwidth, angle=-90]{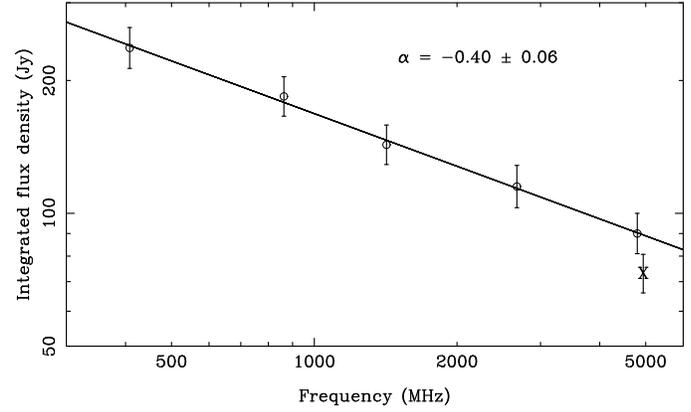}
    \caption{The spectral index was calculated based on flux density values
             determined by \citet{ury+04} and 
             the present $\lambda$6\ cm data. The result of \citet{kb72} is 
             included for comparison and marked by ``X".}
    \label{spec_int}
\end{figure}

We obtained an integrated flux density of the Cygnus Loop at $\lambda$6\ cm 
of $90 \pm 9$\ Jy. To make the result compatible with previous values, 
we have included point sources in the flux integration. The quoted error 
reflects the uncertainties in determining the calibration factors and 
background levels, which adds up in total of about 10\%.
Combining our integrated $\lambda$6\ cm flux density 
with lower frequency data obtained recently by \citet{ury+04}, we derived a spectral 
index of $\alpha=-0.40\pm0.06 $ (Fig.~\ref{spec_int}). This is consistent with 
the result by \citet{ury+04}, who derived $\alpha=-0.42\pm0.06$ up to 
$\lambda$11\ cm. There is no indication of a spectral break in the
frequency range up to $\lambda$6\ cm. The only $\lambda$6\ cm flux density 
of the Cygnus Loop 
previously obtained by \citet{kb72} was $73 \pm 7$~Jy. This flux density 
is lower than our result by 20\% (see Fig.~\ref{spec_int}), which is 
probably caused by missing some diffuse emission due to the lower receiver
sensitivity at that time compared to our present $\lambda$6\ cm system. Their low 
flux density was interpreted as an indication of a spectral steepening.
However, our new $\lambda$6\ cm measurements do not confirm this low flux density. 

\subsubsection{TT-plot analysis}

We made temperature - versus - temperature (TT) plots of the entire source 
at $10\arcmin$ angular resolution between the Urumqi $\lambda$6\ cm and the 
Effelsberg maps at $\lambda$21\ cm and $\lambda$11\ cm to establish a common 
baselevel for these maps. 

\begin{figure}[!htbp]
    \includegraphics[width=0.3\textwidth, angle=-90]{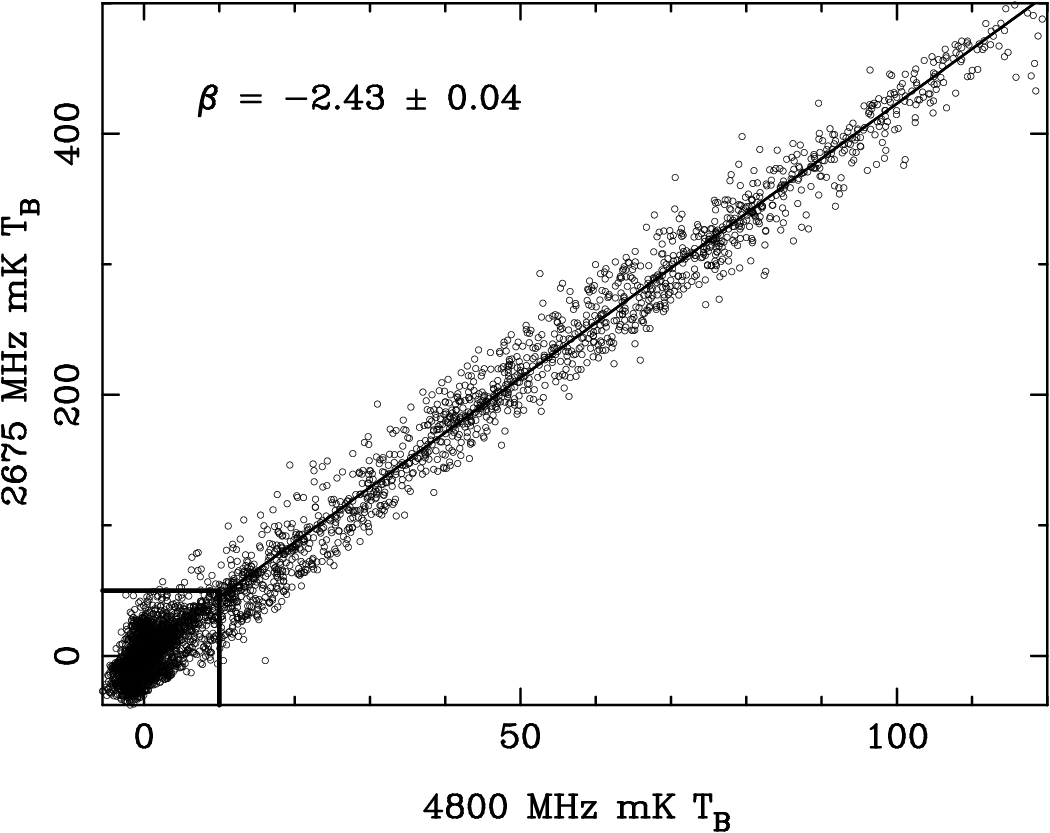}
    \includegraphics[width=0.3\textwidth, angle=-90]{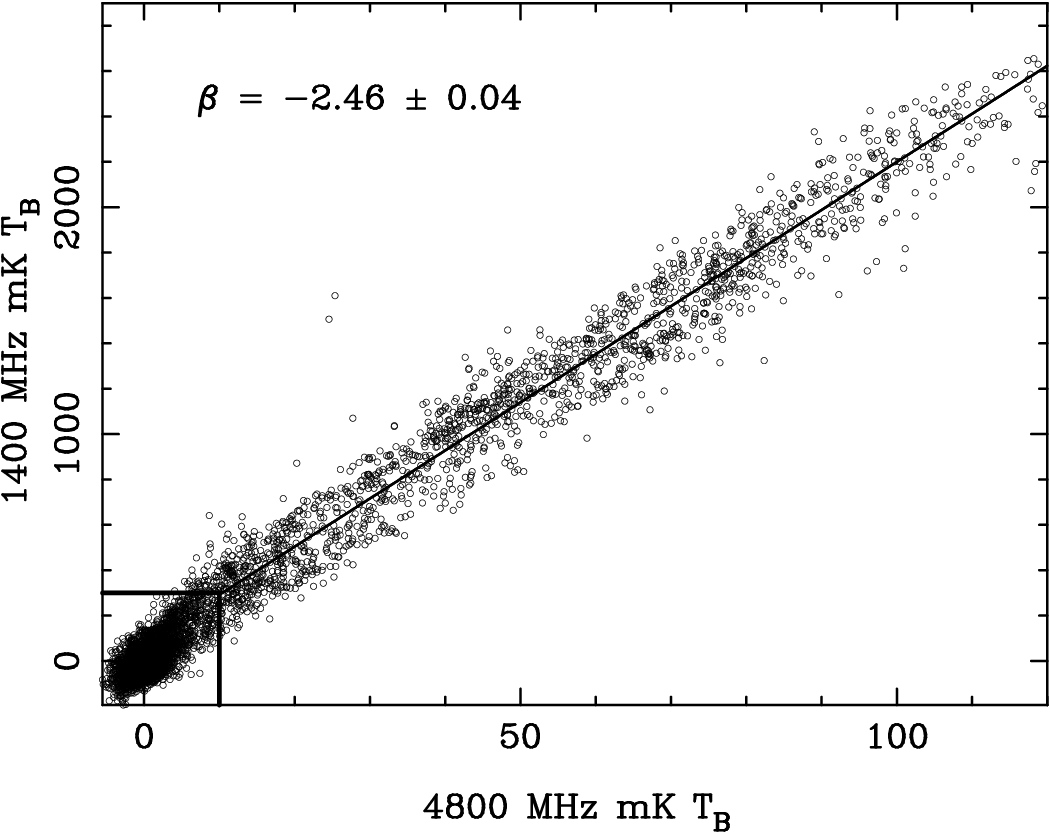}
    \caption{The TT-plots of the $\lambda$6\ cm/$\lambda$11\ cm 
             pair (upper panel) and 
             $\lambda$6\ cm/$\lambda$21\ cm pair (lower panel) are shown.
             The discarded pixels are marked in the small box.}
    \label{tt}
\end{figure}

As the Cygnus Loop is very large, the diffuse Galactic
emission is probably not constant across the entire source. This might 
cause offsets in the TT-plots in particular for low brightness temperatures. 
This effect might also increase the spectral index from TT-plots,
since the spectrum of the Galactic emission is steeper. 
Therefore we have discarded 
all pixels with a brightness temperature below 10\ mK T$_{\rm B}$ at $\lambda$6\ cm, 
50\ mK T$_{\rm B}$ at $\lambda$11\ cm, and 300\ mK T$_{\rm B}$ at 
$\lambda$21\ cm when we fit spectral indices from TT-plots (Fig.~\ref{tt}).

According to the TT-plots, 2\ mK, 8\ mK, and 23\ mK  are 
added to the 4800\ MHz, 2675\ MHz and 1400\ MHz maps. In addition the TT-plot 
yields the spectral index $\beta$ between $\lambda$6 cm/$\lambda$11 cm, 
$\lambda$6 cm/$\lambda$21 cm, and $\lambda$11 cm/$\lambda$21 cm data: $\beta$ =
 $-$2.43$\pm$0.04, $-$2.46$\pm$0.04, and $-$2.47$\pm$0.04, respectively.
The spectral index of the 
brightness temperature $\beta$ and of the flux density $\alpha$ has the 
relation $\alpha=2+\beta$. The average spectral index $\alpha$ 
= $-0.45\pm0.04$ derived from the TT-plot is consistent with 
the spectral index obtained from integrated flux densities. 

\subsubsection{Spectral index map of the Cygnus Loop}

\begin{figure}[!htbp]
\includegraphics[bb=70 140 636 700,width=87mm,clip,angle=-90]{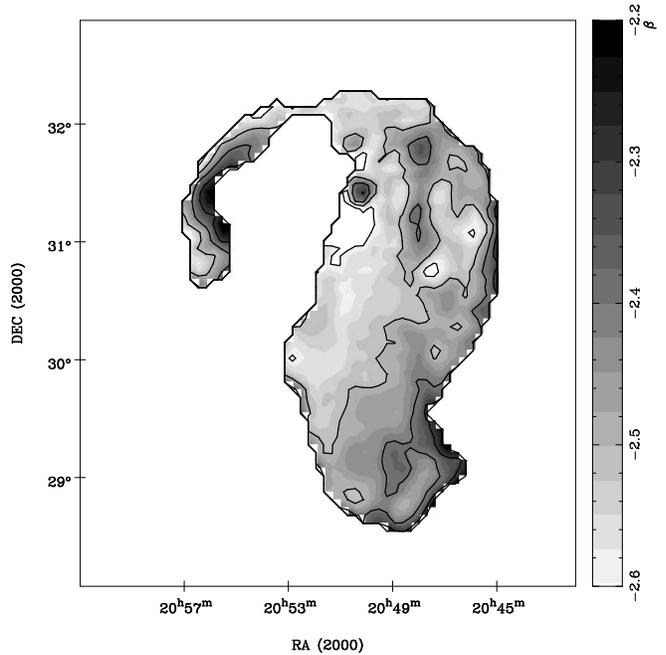}
    \caption{Spectral index map calculated from $\lambda$\ 6cm, 
             $\lambda$11\ cm and $\lambda$\ 21cm maps. All the maps are 
             convolved to the angular resolution of 10\arcmin. The gray 
             scale extends from $-2.6$ (light) to $-2.2$ (dark) and the  
             overlaid contours start at $-2.6$ and run in steps of 0.1.}
    \label{spec_6_11_21}
\end{figure}

Using the maps with corrected background levels we calculated a spectral index 
map  as shown in Fig.~\ref{spec_6_11_21}, where the spectral index 
of each pixel was obtained by linearly fitting the brightness temperatures 
at three freqencies. 
In order to achieve a high signal-to-noise ratio and to exclude the influence
of systematic effects we took the similar cut-off levels as the TT-plots. 
Figure~\ref{spec_6_11_21} shows some distinct regions 
with a higher than average spectral index: the northeastern rim (NGC 6992/5 or NE1/NE2 in 
Fig.~\ref{6cm}) and the 
northwestern region (W in Fig.~\ref{6cm}) both with spectral index of $\alpha \sim -$0.40, 
and the southern part, especially the western edge (S3 inFig.~\ref{6cm}) with spectral index of 
$\alpha \sim -$0.35. Slightly flatter spectra were also reported by ~\citet{lr98} and \citet{ury+04} for these
regions. The spectral index distribution in general shows a gradual steepening from the 
south towards the CL 4 region.
The maximum spectral difference reaches about $\delta\alpha \sim$ 0.3 and is related
 to a decrease of Cygnus Loop's diffuse radio emission (Fig.~\ref{6cm}). It is therefore 
 not clear to what extent the increasing influence of Galactic steep spectrum emission
 in this region  
 is responsible for the gradual spectral steepening or if it is intrinsic to the Cygnus Loop.
This spectral steepening behavior towards the central region with weak diffuse emission 
is similar seen in the spectral index maps of ~\citet{ury+04}.

In the shell of the Cygnus Loop shock acceleration might produce high energy electrons and hence 
strong emission in the presence of a strong magnetic field. 
The intrinsic spectrum of young electrons at the shock may differ from that in the 
diffuse emission region in case energetic particles diffuse away and thus steepen the spectrum,
as synchrotron aging seems unable to explain a steepening.
In regions with weaker magnetic fields the observed emission originates from higher energy electrons.  
The observed slight flattening in the strongest shell regions is qualitatively understandable. 
An alternative explanation based on Galactic magnetic field compression was discussed in
some detail by \citet{lr98}.

It is obvious that a straight spectrum from integrated flux densities 
does not rule out to some extent variations of the spectrum in different regions of a 
SNR or a possible spectral steepening in some areas.
To check this possibility, we compare the spectral index maps $\lambda$6\ cm/$\lambda$11\ cm  
and $\lambda$6\ cm/$\lambda$21\ cm. 
In case the  $\lambda$6\ cm/$\lambda$11\ cm spectral index is smaller than that of 
$\lambda$6\ cm/$\lambda$21\ cm  
a spectral steepening towards higher frequencies is indicated. Otherwise, 
there is a spectral flattening. We found from the two spectral index maps that
in the southern region the spectral indices
agree within $\delta\alpha$ = 0.03.
In the northern part encompassing the central filament, the spectral index of the 
$\lambda$6\ cm/$\lambda$11\ cm pair is smaller by $\delta\alpha$ = 0.1 than the
$\lambda$6\ cm/$\lambda$21\ cm pair. In the NGC 6992/5 
region the spectral index of the $\lambda$6\ cm/$\lambda$11\ cm pair is 
smaller at the edge of the rim and larger near the centre. For the rest of 
the Cygnus Loop the spectral index of the $\lambda$6\ cm/$\lambda$11\ cm pair is 
slightly larger, but never exceeding $\delta\alpha$ = 0.1. These small spectral 
differences  for different regions suggest that 
there is no significant spectral steepening or flattening in the frequency range 
from 1400~MHz to 4800~MHz, even for small regions.
 
We note that a more detailed TT-plot analysis of the Cygnus Loop for a number of 
subfields was made by \citet{green+90}, \citet{lr98} and \citet{ury+04},
where the results from all studies were compared.
Large local spectral curvature as reported by 
~\citet{lr98} was not confirmed by ~\citet{ury+04} 
with high quality data in the frequency range from  408~MHz to 2675~MHz, and 
also not by us for frequencies up to 4800~MHz.

\subsection{Polarization analysis}

\begin{figure*}[!htbp]
\centering
\includegraphics[bb=60 180 642 605,width=0.4\textwidth,clip,angle=-90]{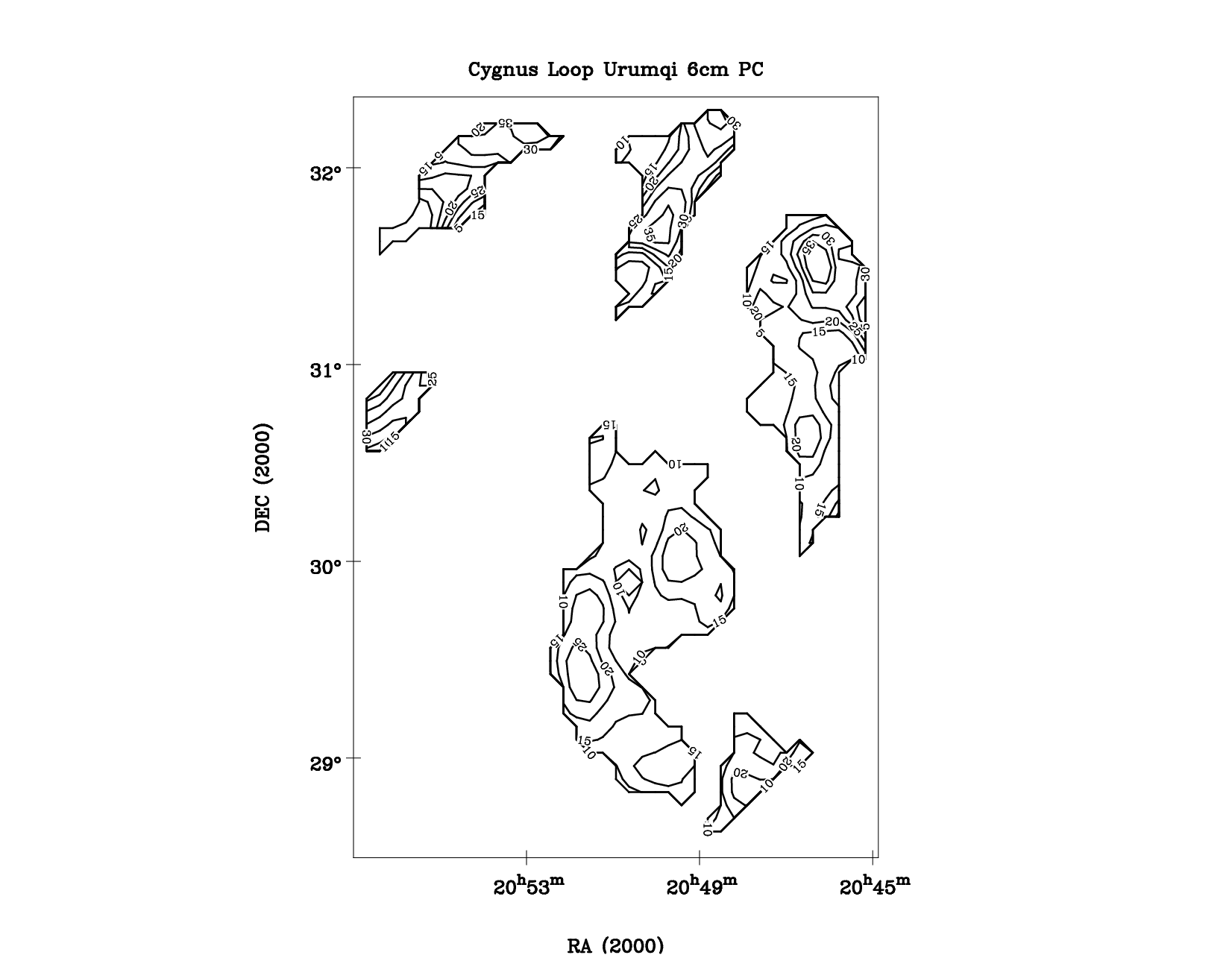}
\includegraphics[bb=60 180 642 605,width=0.4\textwidth,clip,angle=-90]{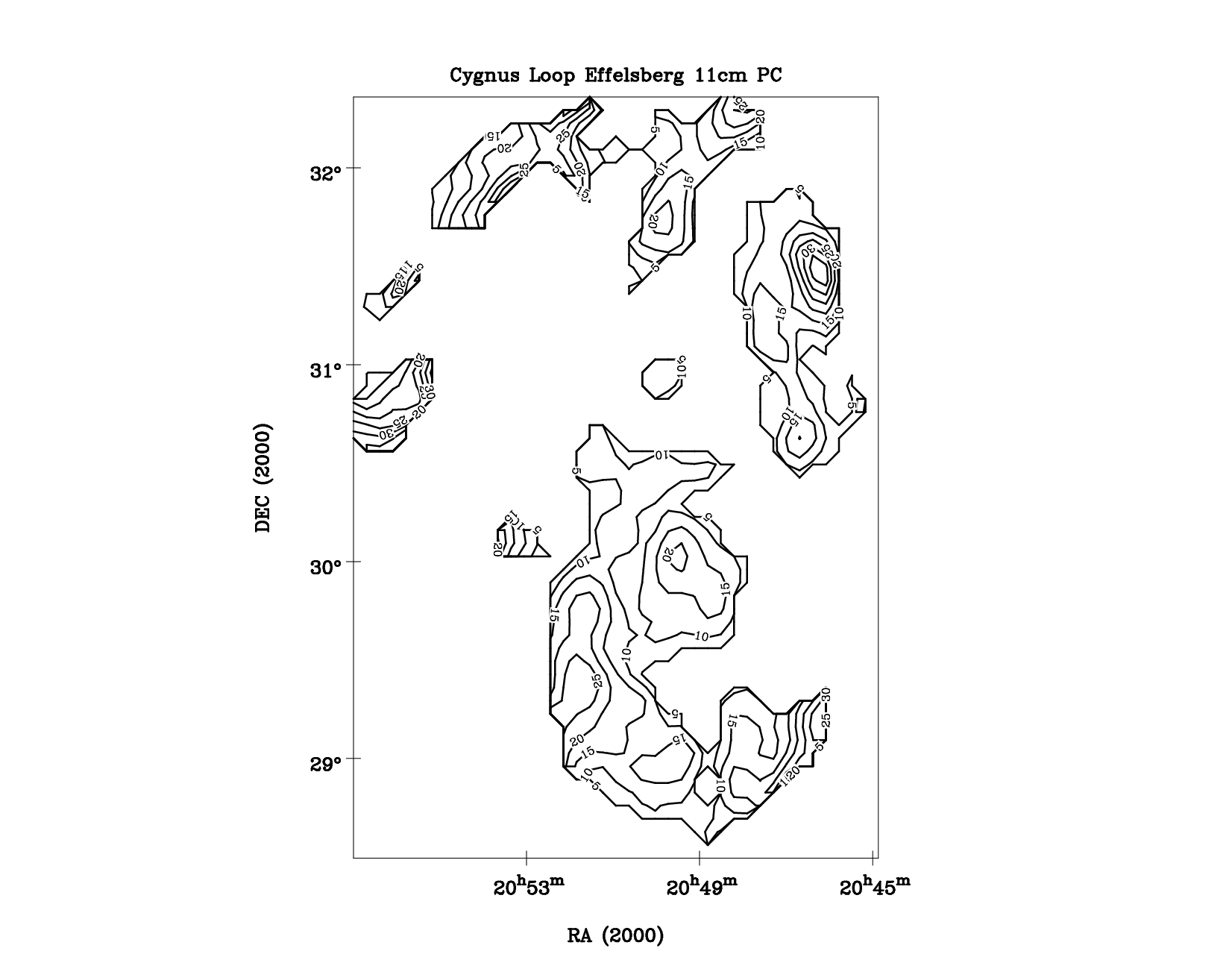}
\includegraphics[bb=60 180 642 605,width=0.4\textwidth,clip,angle=-90]{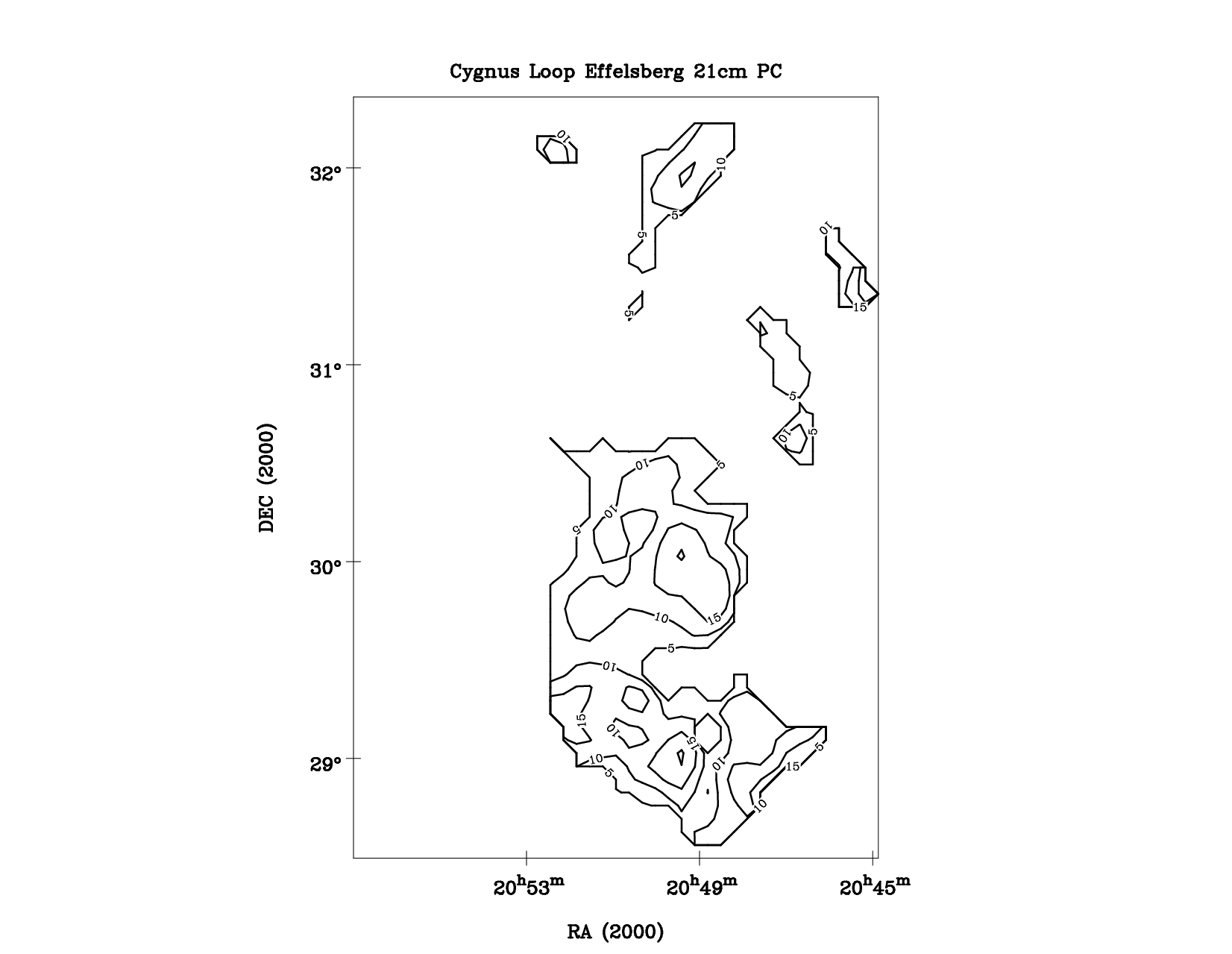}
\caption{Polarization percentage at $\lambda$6\ cm, $\lambda$11\ cm, 
            and $\lambda$21\ cm are shown as contours in left, middle and 
            right panels. All the contours  
            start at a level of  5\% and run in 5\% steps.}\label{pc}
\end{figure*}

The polarization 
intensity maps at $\lambda$11\ cm and $\lambda$6\ cm are very similar in 
morphology (Figs.~\ref{6cm} and \ref{11cm}). The main polarization 
features and their nomenclature are displayed in Fig.~\ref{6cm}. 
The polarization percentage (PC) maps derived for these 
three frequencies are shown in Fig.~\ref{pc}. For total (polarized) intensities below
5x (8x) r.m.s-noise at $\lambda$21\ cm, 5x (10x) r.m.s.-noise at 
$\lambda$11\ cm and 5x (20x) r.m.s.-noise at $\lambda$6\ cm no PC is shown.  

At $\lambda$6\ cm the central filament C in the northern shell shows strong polarization 
with polarization percentages up to 35\% in the northern shell. The NGC 6992/5 
region (NE1 \& NE2 in Fig.~\ref{6cm}) and the NGC 6960 
region (W in Fig.~\ref{6cm}) also show a fair amount of polarization  
up to about 30\%. The southern region consists of three 
patches (Fig.~\ref{6cm}: S1, S2 and S3) and all exhibit considerable 
polarization at a level of about ~20\%. At $\lambda$11\ cm 
the percentage polarization is significantly lower in the north than 
in the south \citep{ury+02}.

A check on the effect of beam depolarization was made by comparing 
the Effelsberg $\lambda$21\ cm map with the higher resolution DRAO $\lambda$21\ cm map 
by \citet{lrb97}. The DRAO map 
shows similar very weak polarization in the northern half of the Cygnus Loop, 
which indicates the presence of strong internal depolarization on scales smaller
than resolved by the $1\arcmin$ beam of their observations.

\subsection{Rotation measure analysis}

The polarization maps of the Cygnus Loop maps at three frequencies in principle allow 
to find out unambiguous rotation measures (RMs). Basically, we linearly fit the 
polarization angles at three frequencies versus the square of the wavelength for 
each pixel from all maps. To gain a high signal to noise 
ratio all pixels with a polarized intensity below $5\times$ the r.m.s-noise 
were not included in the fit.

\begin{figure*}[!htbp]
\includegraphics[bb=48 163 635 622,width=0.6\textwidth,clip,angle=-90]{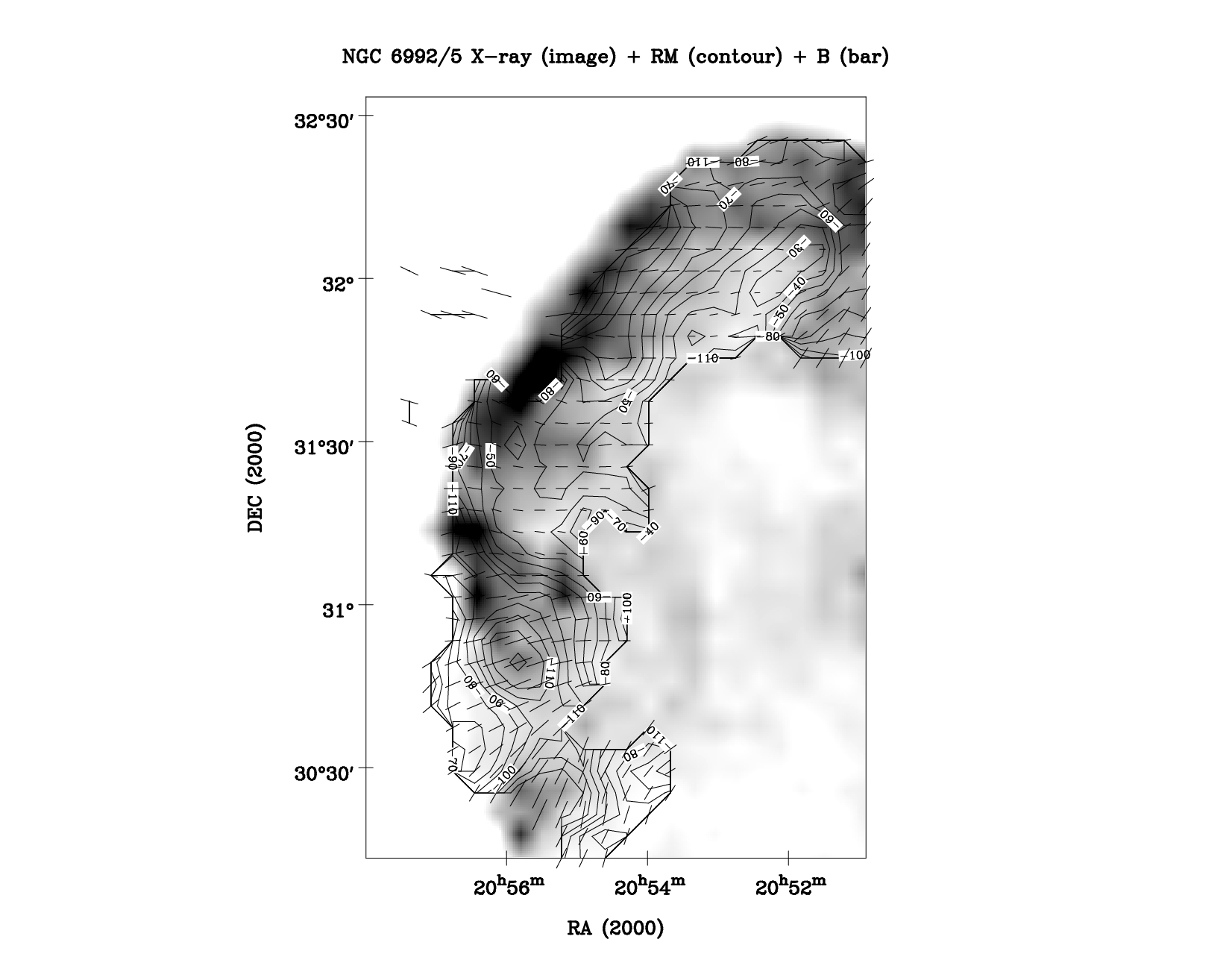}
\includegraphics[bb=55 200 635 600,width=0.6\textwidth,clip,angle=-90]{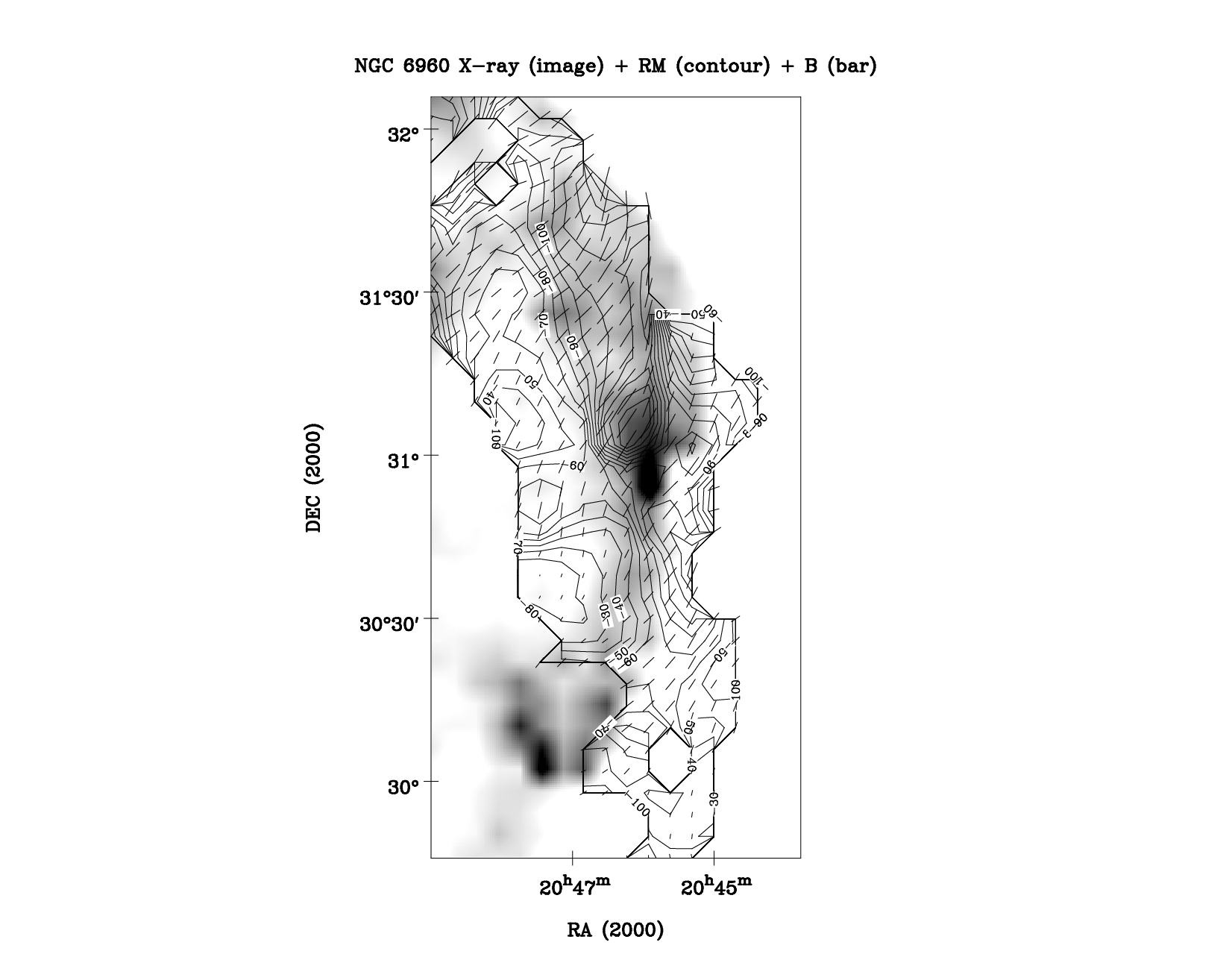}
    \caption{RM maps calculated from the $\lambda\lambda$6\ cm and 11\ cm maps 
             for the NE1/NE2 and the W region are shown 
             in the left and right panels, respectively. 
	     The overlaid grayscale image 
             encodes the 0.25 keV X-ray emission from ROSAT observations \citep{lga+97}. The 
             superimposed bars show the intrinsic orientation of the magnetic field.
     }\label{rm}
\end{figure*}

The average RM in the southern region is $-$21\ rad m$^{-2}$, consistent 
with earlier results by \citet{ury+02}, \citet{lrb97} and \citet{kb72}.
The r.m.s.-scatter ($\sigma_{RM}$) is 7\ rad m$^{-2}$.
Such a RM-value at 4.8~GHz results in a rotation of the polarization angle by 
roughly $4\degr$ with a r.m.s.-scatter of about $1\fdg5$. 
Therefore the $\lambda$6\ cm polarization map quite 
closely displays the intrinsic magnetic field direction in that region (Fig.~\ref{6cm}). 
The maximum {\it systematic} 
uncertainty as infered by the maximum polarization angle deviations of the 
calibration sources ($1\degr$ at $\lambda$11\ cm and at $\lambda$6\ cm), 
respectively, implies a systematic RM error of up to 4~rad m$^{-2}$. 

The average RM in the central filament C in the northern shell of $-28$\ rad 
m$^{-2}$ with $\sigma_{RM}$ of 7\ rad m$^{-2}$ is quite similar to that in
the southern region.

These RM-values can be attributed to the foreground interstellar medium along 
the line of sight in the Galaxy: We assume an electron density $n_e \sim$  
0.02\ cm$^{-3}$, a magnetic field strength along the line-of-sight $B_{||} \sim$ 
3~$\mu$G as typical values for the local interstellar medium and a distance 
$D$ of 540\ pc. The rotation measure RM calculates: 
${\rm RM}=0.81 n_e B_{||} D$, which yields a RM of -26\ rad m$^{-2}$, where the negative 
sign is taken from the general direction of the large scale magnetic field~\citep{han04}. 
We note that the Cygnus Loop distance of 540\ pc was actually derived from the proper 
motion of the optical filament in the NGC 6992/5 region \citep{bsr05}. We use
this distance for all components of the Cygnus Loop. 

Except for filament C, the RMs in the northern part cannot be 
derived unambiguously, since at $\lambda$21\ cm the emission is nearly 
entirely depolarized. For the NGC 6992/5 (NE1/NE2) and NGC 6960 (W) regions 
polarization angles at $\lambda$11\ cm and at $\lambda$6\ cm 
are available. The RM ambiguity
is $\pm$n $\times$ 362\ rad m$^{-2}$. This in principle allows a 
large number of possible RMs, although values of n larger than 1 seem rather unlikely 
in view of the physical size of the objects of a few parsec. 

We show the minimum absolute RMs (n=0) in 
Fig.~\ref{rm}. The average minimum RM is about -73\ rad m$^{-2}$ for NGC 6992/5  
and about -71\ rad m$^{-2}$ for NGC 6960. The r.m.s.-scatter $\sigma_{RM}$ 
is about 31\ rad m$^{-2}$ for both regions, much larger than in the southern part. 
This indicates enhanced fluctuations of the magnetized interstellar medium in the 
shells. The large RM variation in these regions originates interior to the shell, 
which is probably caused by the 
interaction of the blast wave with the cavity wall consisting of a high 
density cloud \citep{lga+97}. Note that the beam of $10\arcmin$ corresponds to 
1.6\ pc at distance of 540\ pc, indicating the presence of 
turbulence in the shock fronts on smaller scales.  

Figure~\ref{rm} shows an intrinsic orientation of the magnetic field in
the shell in a more or less radial direction. This is quite unusual for
an evolved SNR, where a tangential magnetic field is expected. We therefore
analyse also the case of larger RM values (n=$\pm$1).  
Since the RM properties appear similar for NGC 6992/5 and NGC 6960, 
we will focus on the discussion of NGC 6992/5 for the case of large RM values.
The mean RM (n=$\pm$1) is about 290\ rad m$^{-2}$ (or -430\ rad m$^{-2}$). 
Such high values of RM seems in fact possible when we consider the physical conditions
in the shell. The magnetic 
field can be estimated by assuming energy equipartition between the magnetic 
field and electrons and protons~\citep{pac70}:

\begin{equation}
B_{\min}  = C \cdot V\, \mbox{}^{-2/7} \cdot L\, \mbox{}^{2/7}
\end{equation}

We derive a $\lambda$6\ cm flux density of 11\ Jy by integrating the NE1 
and NE2 region and calculate the luminosity L for a spectral index of $\alpha$ = $-0.40$.  
For the radiating volume V we estimate about 400 pc$^3$. C is a constant 
\citep{pac70}.

We obtain an estimate of the magnetic field of 47$\mu$G and the line-of-sight 
component of about 33 $\mu$G by multiplying the factor of 1/$\sqrt{2}$.
The typical electron density derived from X-ray observations is 
a few cm$^{-3}$ with a temperature of 
about 0.2 keV from ASCA observations~\citep{mtpk94}. A density of about 
1 cm$^{-3}$ in the shell with a similar temperature was derived from ROSAT 
observations \citep{luinprep}. The postshock electron 
density derived from the optical filament observation is 10-100~cm$^{-3}$ 
(O{\scriptsize VI} doublet~\citep{lbv+92} or S{\scriptsize II}~\citep{pfr+02}),
much larger than that from the X-ray data, although the X-ray emission originates from
diffuse gas. With a temperature of 0.1~keV and an electron 
density of 1~cm$^{-3}$ the pressure is $1\times10^{-10}$ 
erg cm$^{-3}$. For a magnetic field of 47~$\mu$G the 
pressure is $0.87\times10^{-10}$ erg cm$^{-3}$, which means a balance 
between the gas and the magnetic pressure. 

Taking the values of the electron density and the magnetic field estimated above
and the size of the rim as about 4.8~pc, we calculate an absolute RM of 
128\ rad m$^{-2}$. This RM value shall be taken as a strict lower limit, because 
the magnetic field can be underestimated by a factor of several due to the 
unclear filling factor and the electron density is locally definitely much higher
than 1~cm$^{-3}$.

\begin{figure*}[!htbp]
\includegraphics[bb=64 167 630 590,width=0.6\textwidth,clip,angle=-90]{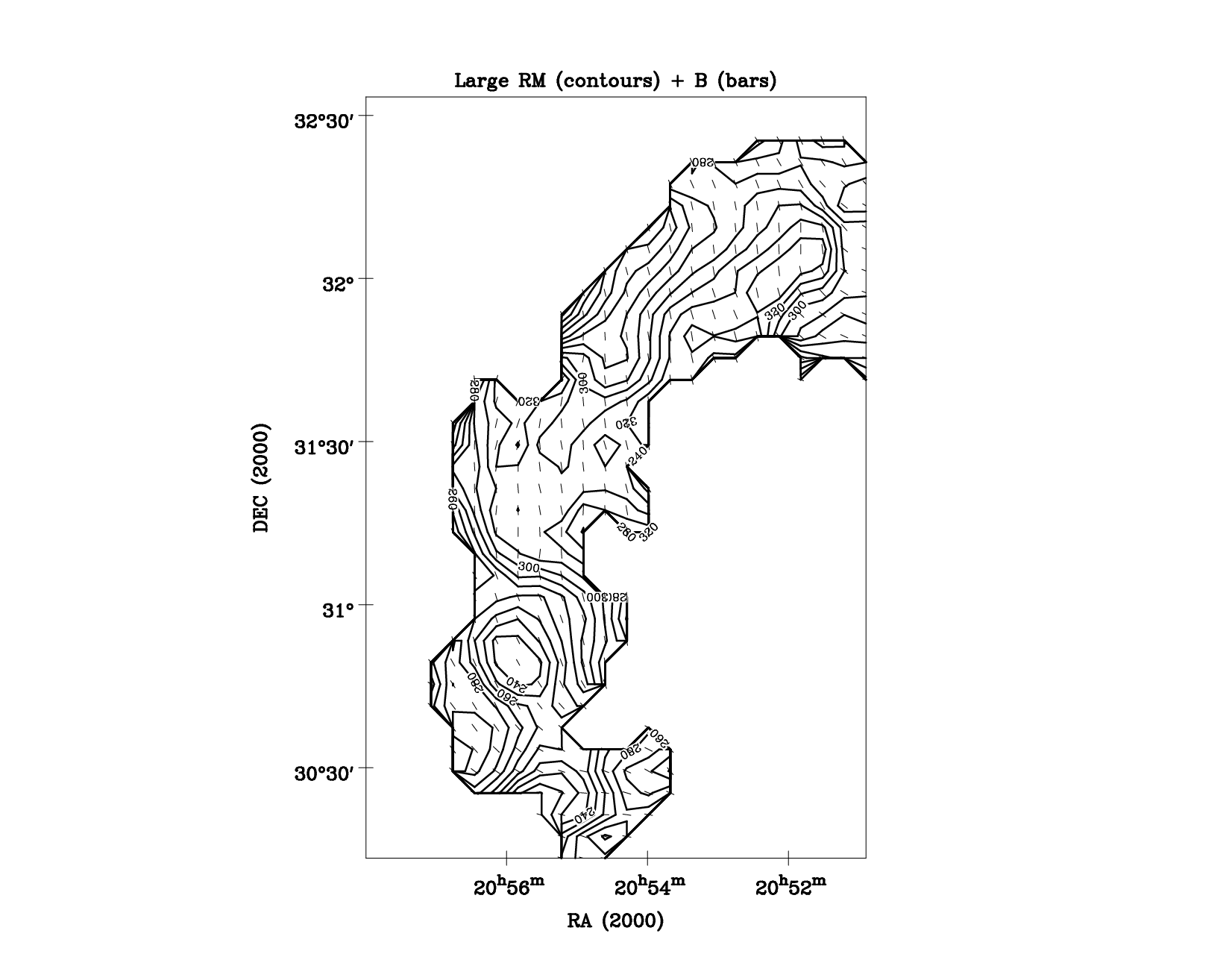}
\includegraphics[bb=59 220 634 544,width=0.6\textwidth,clip,angle=-90]{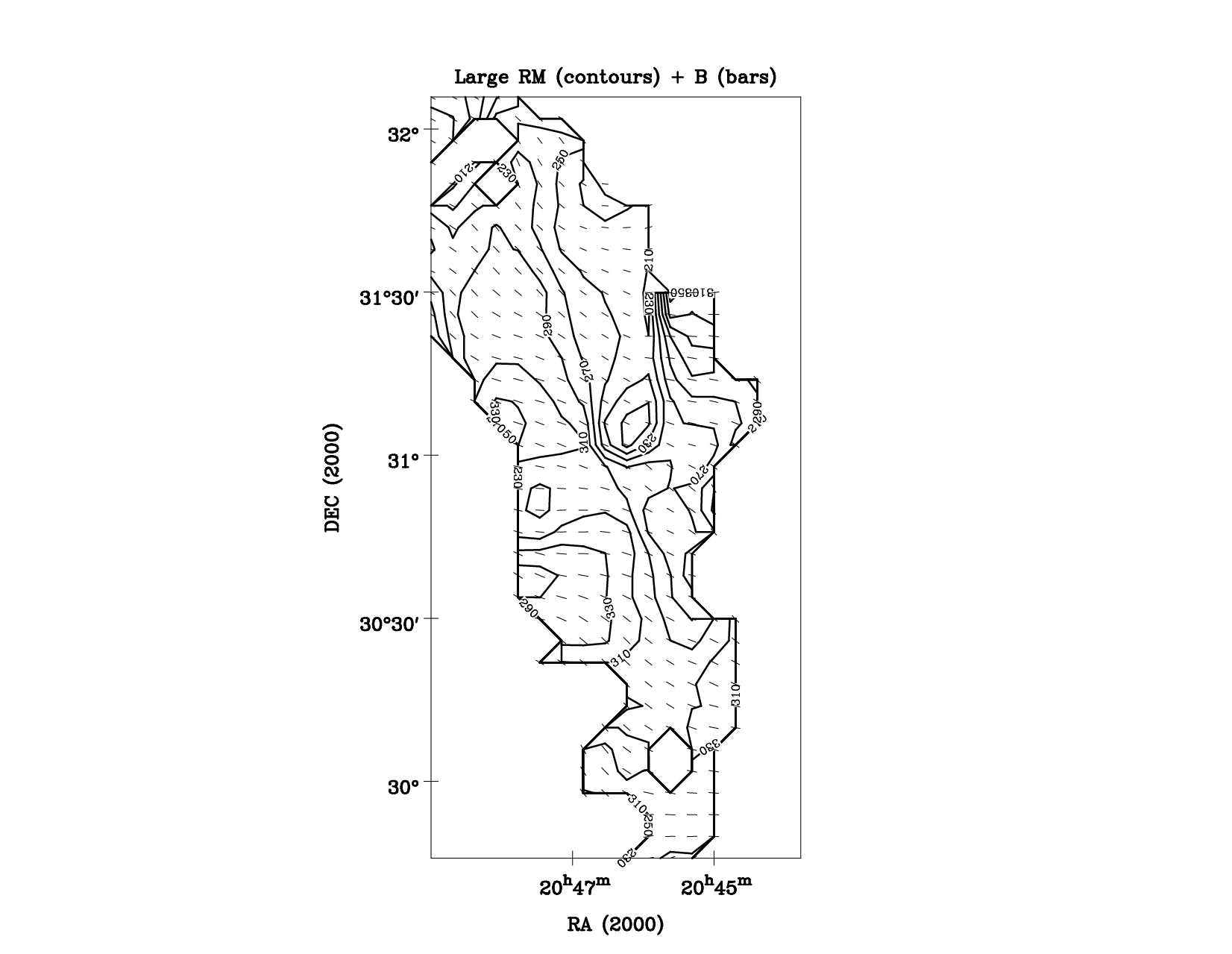}
  \caption{Rotation Measure in contours and the intrinsic magnetic field 
           in bars for the large RM case is shown. 
}\label{rm_large}
\end{figure*}

The large RMs rotate the $\lambda$6\ cm polarization angles by about $60\degr$ and
the intrinsic magnetic field direction is in this case along the shell 
especially for the NE1 knots and for NGC~6960 (Fig.~\ref{rm_large}), but not for NE2.  
The NE1/NE2 shell is  
considerably evolved and has already entered the radiative phase~\citep{dcl+00}. At this 
stage, the frozen-in interstellar magnetic field is compressed in the 
shell~\citep{laa62} and is expected to be tangential to the shock 
front~\citep{fr04} as observed. 

We conclude that from our estimates large RM values like 290 (-430)~rad m$^{-2}$ are quite feasible
and in agreement with a basically tangential magnetic field in the shell. The minimm RM 
value of -73 ~rad m$^{-2}$ can not be ruled out, but requires a more radial field configuartion,
which is typical for rather young SNRs in free expansion. A problem with
RMs calculated from two frequencies is that there is no proof of the validity of the 
$\lambda^{2}$ dependence to extrapolate for intrinsic values. This happens in case 
the observed polarization at the two frequencies does not originate in the same volume.
For a proof polarization observations higher than 5~GHz are needed. 

\subsection{Depolarization analysis}

 The degree of depolarization
$DP$ is defined as $DP=PC_o/PC_i$, where $PC_o$ is the observed
polarization percentage and $PC_{i}$ is the intrinsic polarization
percentage. For smaller $DP$ the depolarization is more reduced. The caveat
here is that $DP$ is wavelength-dependent and
$PC_i$ is assumed to be constant for different wavelengths. 

There are several mechanisms which could cause depolarization~\citep{bur66,tri91,sbs+98}. 
One of these mechanisms is {\it bandwidth depolarization}, which
occurs when the polarization angles are rotated by a different amount across the band
and hence the observed polarization is reduced. 
{\it Bandwidth depolarization} can be written as
$DP={\rm sinc}(2RM\lambda^2\frac{\Delta\nu}{\nu})$, where $\Delta\nu$ is
the bandwidth. Another mechanism is {\it external depolarization}. This is severe
in the case of large RM fluctuations ($\sigma_{RM}$) in the foreground medium.
Different regions of background polarization are rotated to a
different extent and reduce the summed up polarization signal.
The degree of depolarization is directly related with $\sigma_{RM}$ as
$DP=\exp(-2\sigma_{RM}^2\lambda^4)$. Finally there is {\it internal depolarization}
which occurs in the emission region. The average of the polarization from
different depth and hence with different orientations within the source results 
in depolarization. In this mechanism $DP$ is given by
$DP=|\frac{1-\exp(-S)}{S}|$, where
$S=2\sigma_{RM}^2\lambda^4-2i\lambda^2\mathcal R$.
Here $\mathcal R$ is the RM through the entire source,
${\mathcal R}=2RM$~\citep{sbs+98}. When the variation of RM is very small,
the degree of depolarization reduces to $DP={\rm sinc}\mathcal R\lambda^2$.

For the southern region the effect of the {\it bandwidth depolarization} is
negligible.  
With a bandwidth of 20~MHz at $\lambda$21\ cm, 40~MHz at $\lambda$11\ cm, 
600~MHz at $\lambda$6\ cm and a RM of $-$21\ rad\ m$^{-2}$,
a $DP$ of $\sim 1$ is obtained for all the cases. Internal depolarization
seems very small since the RM is mainly of external origin. 
{\it External depolarization} does not cause large
depolarization due to the small variation of RM. Therefore
$PC_{6cm}\approx PC_{11cm}\approx PC_{21cm}\approx PC_{i}$ is predicted. In fact, we measure
similar PCs (Fig.~\ref{pc}) at $\lambda$6\ cm and
$\lambda$11\ cm, and a slightly smaller PC at
$\lambda$21\ cm map. All measurements are consistent with the expectations.

For the northern part the situation
is more complex. Polarized intensities were not detected
for NE1 and NE2 and marginally detected for W at $\lambda$21\ cm, which means
that $DP_{\rm 21cm}$ is nearly zero at these regions. The average PC is 
measured to be about 27\%, 28\% and 20\% for NE1, NE2 and
W regions at $\lambda$6\ cm and 23\%, 23\% and 15\% at $\lambda$11\ cm.
Here, we investigate the two possibilities for the RM, the
small RM case ($-70$\ rad m$^{-2}$) and the large RM case (290 ($-$430)\ rad
m$^{-2}$) to see which can account for the observed depolarization. For both
cases the {\it bandwidth depolarization} is very small.
Internal depolarization, therefore, plays an important role
since the large RM and RM variation are caused by a medium interior to the SNR.
\citet{lrb97} found a correlation between
X-ray emission and depolarization, which also indicates the presence of
{\it internal depolarization}. According to that both the large
RM case and the small RM case can result in the nearly entire depolarization
at $\lambda$21\ cm. A RM of $-$430\ rad\ m$^{-2}$, however, can be excluded because
this high value of RM makes the polarization even at $\lambda$6\ cm
virtually vanish.
The small RM can produce $DP$ of $\sim$94\%, which indicates that the polarization
at $\lambda$6\ cm is nearly intrinsic. But a $DP$ of $\sim$50\% is
then predicted for $\lambda$11\ cm. This means the observed PC at $\lambda$11\ cm 
is larger than expected.
It could well be that the observed polarization originates from different
regions along the line of sight at $\lambda$11\ cm and $\lambda$6\ cm, where
it might partly come from larger distances than at $\lambda$11\ cm.
A large RM of 290\ rad m$^{-2}$, however, causes about 33\% of $DP$ at $\lambda$6\ cm,
implying an intrinsic polarization of 80\% for the NE1 and NE2 regions and
60\% for the W region, which is critical although still feasible. The expected
PC is also smaller than the observed PC at $\lambda$11\ cm
as for the small RM case.
We conclude that we cannot infer from the depolarization analysis
between the case of large RM or small RM. A definite answer will be given by
another high frequency observation or multi-channel narrow band polarimetry.

The central filament shows 
little depolarization and is salient at all the three wavelengths. 
This filament is totally different from the other parts in the north in 
polarization property, which results from its location likely on the surface
 of the expanding shells. 

\section{Remarks on two SNRs senario}

\citet{ury+02} proposed that the Cygnus Loop consists of two
SNRs: G74.3$-$8.4 and G72.9$-$9.0, where in the centre of G72.9$-$9.0
the anomalous X-ray source AX J2049.6+2939 is located. It is most likely 
a neutron star \citep{mot+01} left from the supernova explosion 
of a 11-20 M\sun ~~progenitor star~\citep{lea04}. 
From the two polarization maps at $\lambda$6\ cm (Fig.~\ref{6cm}) and 
$\lambda$11\ cm (Fig.~\ref{11cm}), 
we can clearly identify two polarized shells corresponding to G74.3$-$8.4 
and G72.9$-$9.0 with a partial overlap. The 
polarization properties of these two shells are different. In the southern 
shell the magnetic field follows the shell conforming to the standard 
 picture of a middle-aged SNR in the adiabatic phase. In the northern part 
 the magnetic field orientation seems to be tangential as well, but the shell structure
 is more  
irregularly shaped by the blast wave-cloud interaction. The northern shell 
is totally depolarized at $\lambda$21\ cm in contrast to the southern shell. 
Differences between the northern and southern shell are also evident in their
X-ray and optical appearance.
All the facts above strongly support the two-SNR scenario proposed by \citet{ury+02}.

The relation between these two SNRs is not quite clear as already mentioned 
by \citet{ury+02}. From our RM analysis the two SNRs are at about the same distance.
In the region of overlap there are two regions with enhanced X-ray 
emission at the western and estern side \citep{al99,ury+02} 
indicating some interaction between the two SNRs.
The spectral index distibution (Fig.~\ref{spec_6_11_21}) shows a weak gradient, but no
distinct change.
Surprisingly the X-ray emission does not show any enhancement in the region of overlap 
 \citep{lga+97} as one might expect from the interaction of two shock waves. As already
 noted by \citet{ury+02} this is not unique as such an enhancement is also not
 observed from the interacting SNRs DEM L316 in the LMC \citep{wil97}. A more 
 advanced model and further
observations of the Cygnus Loop are needed for clarification. 

\section{Conclusion}

We present a new sensitive total intensity and polarization map of the Cygnus Loop 
at $\lambda$6\ cm, which is the shortest wavelength where a complete map including
polarization has been obtained so far. Our highly stable receiver enables us to 
accurately measure the faint extended emission of the source, which leads to a flux 
density increase by 20\% compared to older data. Our spectral index of $\alpha=-0.40$ 
for the wavelength range $\lambda$21\ cm to $\lambda$6\ cm is consistent with that 
found by \citet{ury+04} for longer wavelength. 
This rules out a spectral break in the frequency range between 1.4~GHz to 4.8~GHz. 
The spectral index maps between different wavelength pairs do not show any spectrum 
curvature. Some small spatial variations of the spectral index is visible on the  
combined spectral index map (Fig.~\ref{spec_6_11_21}).
The polarization maps support the idea that the Cygnus Loop consists of two SNRs. 
We derived a RM of $-21$\ rad m$^{-2}$ towards the southern SNR and a RM of $-28$\ rad m$^{-2}$
for filament C. These RM-values are explained by the Galactic foreground emission.   
The turbulent magnetized medium interior to the northern shell caused by the 
interaction of a blast wave with a cloud causes total depolarization 
at $\lambda$21\ cm. The RM of the northern 
part is based on $\lambda$11\ cm and $\lambda$6\ cm observations. Both a large mean RM of about 
290 rad m$^{-1}$ or a small mean RM of about $-$70 rad m$^{-2}$ seem possible.
The larger RM implies a tangential magnetic field for most sections of the shell
as expected for an evolved SNR.

\begin{acknowledgements}
The $\lambda$6\ cm data were obtained with the receiver system from MPIfR 
mounted at the Nanshan 25--m telescope at the Urumqi Observatory of NAOC. 
We thank the staff of the Urumqi Observatory of NAOC for the great 
assistance during the installation of the receiver and the observations. 
In particular we like to thank Mr. 
Otmar Lochner for the construction of the $\lambda$6\ cm system and its 
installation and  Mr. M. Z. Chen and Mr. J. Ma for their deep engagement 
during the installation of the receiver and their maintenance efforts to 
preserve its excellent performance. We are very grateful to Dr. Peter M\"uller 
for development or adaptation of software needed to make mapping observations and 
data reduction possible at the Urumqi site. The MPG and the NAOC supported the 
construction of the Urumqi $\lambda$6\ cm receiving system by special funds. 
The research work of XHS and JLH was supported by the National Natural Science
foundation of China (10473015) and by the Partner group of MPIfR at NAOC. 
Great support for many bilateral visits from the exchange program between MPG and CAS 
is appreciated.
Finally we like to thank Nancy Levenson for providing the ROSAT soft X-ray image
and Ernst F\"urst for critical reading of the manuscript.
\end{acknowledgements}

\bibliographystyle{aa}

\begin{thebibliography}{}
\bibitem[Aschenbach \& Leahy(1999)]{al99}
Aschenbach, B., \& Leahy, D. A. 1999, A\&A, 341, 602
\bibitem[Blair, Sankrit \& Raymond(2005)]{bsr05}
Blair, W. P., Sankrit, R., \& Raymond, J. C. 2005, AJ, 129, 2268
\bibitem[Burn(1966)]{bur66}
Burn, B. J. 1966, MNRAS, 133, 67
\bibitem[Condon et al.(1998)]{ccg+98}
Condon, J. J., Cotton, W. D., Greisen, E. W., Yin, Q. F., Perley, R. A., Taylor, G. B., \& Broderick, J. J. 1998, AJ, 115, 1693.
\bibitem[Danforth et al.(2000)]{dcl+00}
Danforth, C. W., Cornett, R. H., Levenson, N. A., Blair, W. P., \& Stecher T. P. 2000, AJ, 119, 2319
\bibitem[DeNoyer(1974)]{den74}
DeNoyer, L. K. 1974, AJ, 79, 1253
\bibitem[Desai \& Fey(2001)]{df01}
Desai, K. M., \& Fey, A. L. 2001, ApJS, 133, 395
\bibitem[Emerson \& Gr\"ave(1988)]{eg88}
Emerson, D. T., \& Gr\"ave, R. 1988, A\&A, 190, 353
\bibitem[F\"urst \& Reich(1986)]{fr86}
F\"urst, E., \& Reich, W. 1986, A\&A, 163, 185
\bibitem[F\"urst \& Reich(2004)]{fr04}
F\"urst, E., \& Reich, W. 2004, in "The Magnetized Interstellar Medium", 
eds. B. Uyan{\i}ker, W. Reich, \& R. Wielebinski, Copernicus GmbH, p. 141
\bibitem[Goss, van Gorkom \& Shaffer(1979)]{goss+79}
Goss, W. M., van Gorkom, J., \& Shaffer, D. B. 1979, A\&A, 73, L17 
\bibitem[Green (1990)]{green+90}
Green, D. A. 1990, AJ, 100, 1927
\bibitem[Han(2004)]{han04}
Han, J. L. 2004, in "The Magnetized Interstellar Medium",
eds. B. Uyan{\i}ker, W. Reich, \& R. Wielebinski, Copernicus GmbH, p. 3
\bibitem[Haslam(1974)]{has74}
Haslam, C. G. T. 1974, A\&AS, 15, 333
\bibitem[Hatchell(2003)]{hat03}
Hatchell, J. 2003, APEX-IFD-MPI-0002
\bibitem[Keen et al.(1973)]{kwh+73}
Keen, N. J., Wilson, W. E., Haslam, C. G. T., Graham, D. A., \& Thomasson, P. 1973, A\&A, 28, 197
\bibitem[Kundu \& Becker(1972)]{kb72}
Kundu, M. R., \& Becker, R. H. 1972, AJ, 77, 459
\bibitem[Leahy(2002)]{lea02}
Leahy, D. A. 2002, AJ, 123, 2689
\bibitem[Leahy(2004)]{lea04}
Leahy, D. A. 2004, MNRAS, 351, 385
\bibitem[Leahy, Roger \& Ballantyne(1997)]{lrb97}
Leahy, D. A., Roger, R. S., \& Ballantyne, D. 1997, AJ, 114, 2081
\bibitem[Leahy \& Roger(1998)]{lr98}
Leahy, D. A., \& Roger, R. S. 1998, ApJ, 505, 784
\bibitem[Levenson et al.(1997)]{lga+97}
Levenson, N. A., Graham, J. R., Aschenbach, B., et al. 1997, ApJ, 484, 304
\bibitem[Long et al.(1992)]{lbv+92}
Long, K. S., Blair, W. P., Vancura O., et al. 1992, ApJ, 400, 214
\bibitem[Lu \& Aschenbach (2005)]{luinprep}
Lu, F. J., \& Aschenbach, B. 2005, A\&A, to be submitted
\bibitem[Miyata et al.(1994)]{mtpk94}
Miyata, E., Tsunemi, H., Pisarski, R., \& Kissel, S. 1994, PASJ, 46, L101
\bibitem[Miyata et al.(2001)]{mot+01}
Miyata, E., Ohta, K., Torii, K., et al. 2001, ApJ, 550, 1023
\bibitem[Pacholczyck (1970)]{pac70}
Pacholczyck, A. G. 1970, Radio Astrophysics, W.H. Freemann and Company, San Francisco, p. 171
\bibitem[Patnaude et al.(2002)]{pfr+02}
Patnaude, D. J., Fesen, R. A., Raymond, J. C., et al. 2002, AJ, 124, 2118
\bibitem[Sofue \& Reich(1979)]{sr79}
Sofue, Y., \& Reich, W. 1979, A\&AS, 38, 251
\bibitem[Sokoloff et al.(1998)]{sbs+98}
Sokoloff, D. D., Bykov, A. A., Shukurov, A., Berkhuijsen, E. M., Beck, R., 
\& Poezd, A. D. 1998, MNRAS, 299, 189
\bibitem[Tenorio-Tagle, R\'ozyczka, \& Yorke(1985)]{ten85} 
Tenorio-Tagle, G., R\'ozyczka, M., \& Yorke, H. W. 1985, A\&A, 148, 52
\bibitem[Tribble(1991)]{tri91}
Tribble, P. C. 1991, MNRAS, 250, 726
\bibitem[Uyan{\i}ker et al.(1999)]{bu+99}
Uyan{\i}ker, B., F\"urst, E., Reich, W., Reich, P., \& Wielebinksi, R. 1999, A\&AS, 138, 31
\bibitem[Uyan{\i}ker et al.(2002)]{ury+02}
Uyan{\i}ker, B., Reich, W., Yar, A., Kothes, R., \& F\"urst, E. 2002, A\&A, 389, L61
\bibitem[Uyan{\i}ker et al.(2004)]{ury+04}
Uyan{\i}ker, B., Reich, W., Yar, A., \& F\"urst, E. 2004, A\&A, 426, 909
\bibitem[van der Laan(1962)]{laa62}
van der Laan, H. 1962, MNRAS, 124, 179
\bibitem[Wang et al. (2001)]{wang+01}
Wang, N., Manchester, R. N., Zhang, J., Wu, X. J., Yusup, A., Lyne, A. G., Cheng, K. S., \& Chen. M. Z.
2001, MNRAS, 328, 855
\bibitem[Wang et al. (2003)]{wang+03}
Wang, N., Manchester, R. N., Zhang, J., Wu, X. \& Esamudin, A.
2003, Acta Astronomica Sinica, 44, 207
\bibitem[Wielebinski et al. (2002)]{rw+02}
Wielebinski, R., Lochner, O., Reich, W., \& Mattes, H. 2002, in "Astrophysical Polarized Backgrounds",
eds. S. Cecchini et al., AIP Conference Proceedings, Vol. 609, p. 291
\bibitem[Williams et al. (1997)]{wil97}
Williams, R. M., Chu Y.-H., Dickel, J. R., Beyer, R., Petre, R., Smith, R. C., \& Milne, D. K.
1997, ApJ, 480, 618
\end{thebibliography}
\end{document}